\documentclass[prd,11pt,nofootinbib,preprint]{revtex4}

\usepackage[T1]{fontenc}
\usepackage{amsmath,amssymb}
\usepackage{epsfig}
\usepackage{url}
\usepackage{subfigure}
\usepackage{graphicx}
\usepackage{grffile}
\usepackage[usenames,dvipsnames]{color}
\usepackage{slashed}
\usepackage{array}
\usepackage{multirow}
\usepackage[colorlinks,citecolor=blue]{hyperref}
\usepackage{color}
\usepackage{cancel}
\usepackage{gensymb}

\def\a{\alpha}

\def\b{\beta}

\def\l{\lambda}
\def\m{\mu}

\def\D{\Delta}

\def\q2 {q^2}

\def\t {\theta}

\def\a {\alpha}
\def\b {\beta}

\def\l {\lambda}

\def\bar {\overline}

\def\be {\begin{equation}}
\def\ee {\end{equation}}
\def\beq {\begin{equation}}
\def\eeq {\end{equation}}

\newcommand{\besub}{\begin{subequations}}
\newcommand{\eesub}{\end{subequations}}
\newcommand{\bea}{\begin{eqnarray}}
\newcommand{\eea}{\end{eqnarray}}

\def\beq{\begin{equation}}
\def\eeq{\end{equation}}
\def\barr{\begin{array}}
\def\earr{\end{array}}

\begin{document}
\title{The muon $g-2$ and $W$-mass anomalies explained and the electroweak vacuum stabilised by extending the minimal Type-II seesaw}

\author{Nabarun Chakrabarty}
\email{nabarunc@iitk.ac.in, chakrabartynabarun@gmail.com}
\affiliation{Department of Physics, Indian Institute of Technology Kanpur,
Kanpur, Uttar Pradesh 208016, India}

\begin{abstract} 
The recent precise measurement of the $W$-mass by the CDF II collaboration is indicative of new physics beyond the Standard Model. On the other hand, a resolution of the longstanding muon $g-2$ anomaly also calls for additional dynamics. In this work, we accommodate the two aforementioned anomalies in an extension of the minimal Type-II seesaw model. That is, the minimal Type-II model is augmented with an additional doubly charged scalar and vector leptons. While a chirality-flip of the vector leptons can predict the observed value of muon $g-2$, the value of the recently reported $W$-mass can also be simultaneously achieved
through the oblique parameters of the model. In addition, we further show that the parameter region allowing for the simultaneous resolution of the two anomalies complies with the neutrino mass data, lepton flavour violation and electroweak vacuum stability up to the Planck scale. 
\end{abstract} 
\maketitle

\section{Introduction}\label{intro}

The discovery of the Higgs boson of mass 125 GeV~\cite{Chatrchyan:2012xdj,Aad:2012tfa} at the Large Hadron Collider (LHC)
completes the particle spectrum of the Standard Model (SM). 
Moreover, the interactions of the boson with SM fermions and gauge bosons are increasingly in agreement with the corresponding SM values. Despite this success, certain pressing inconsistencies within the SM necessitate additional dynamics beyond-the-SM (BSM). That the SM alone cannot stabilise the electroweak (EW) vacuum up to the Planck scale is one such theoretical shortcoming~\cite{Degrassi:2012ry,Buttazzo:2013uya,Zoller:2014cka,EliasMiro:2011aa,Isidori:2001bm}. However, additional bosonic degrees of freedom over and above the SM ones can potentially offset this destabilising effect coming from the t-quark (see the references in \cite{Swiezewska:2016rrp}). The remedy in this case lies in introducing additional scalar degrees of freedom that can potentially counter the effect from the t-quark and stabilise the vacuum till the Planck scale. On the experimental front, for instance, the SM fails to account for the observed non-zero neutrino masses and their mixings. However,
appropriately augmenting the SM by additional fields can lead to a non-zero neutrino mass via the seesaw mechanism.
Of these, the popular Type-II seesaw~\cite{PhysRevD.22.2227,Magg:1980ut,Lazarides:1980nt} employs a complex scalar $SU(2)_L$ triplet and is also known to be attractive from the perspective of baryogenesis and collider signatures. It has also been shown to alleviate the vacuum instability problem~\cite{Chun:2012jw,Dev:2013ff,Chakraborty:2014xqa}.

In addition, certain fresh experimental results 
over the past few years have reinforced the claims of additional dynamics beyond the SM. First, the long-standing discrepancy in the muon anomalous magnetic moment reported by
Brookhaven E821~\cite{Muong-2:2006rrc} has now been confirmed by Fermilab E989 through its "MUON G-2" experiment~\cite{Muong-2:2021ojo,Muong-2:2021vma}. The combined result is quoted as 
\bea
\Delta a_\mu = (2.51 \pm 0.59) \times 10^{-9},
\eea
which is $4.2 \sigma$ away from the SM prediction. Secondly, a recent measurement of the $W$-boson mass by the CDF collaboration sets the value at~\cite{cdfII:2022} 
\bea
M_W = 80.4335~\text{GeV} \pm 6.4~\text{MeV} (\text{stat}) \pm 6.9~\text{MeV}(\text{sys}),
\eea
an apparent $7.2\sigma$ disagreement
with the SM value, i.e., $M_W = 80.357 \pm 
4~\text{MeV}(\text{stat}) \pm 4$ MeV(sys). If such a discrepancy persists in other experimental data, it must indicate presence of new physics (NP) encoded through the oblique parameters~\cite{PhysRevD.46.381,PhysRevLett.65.2967}. In fact, the announcement of the $W$-mass has spurred a series of investigations, each invoking some NP scenario, some of the earliest ones being~\cite{Strumia:2022qkt,deBlas:2022hdk,Paul:2022dds,Gu:2022htv,Asadi:2022xiy,Endo:2022kiw,Balkin:2022glu}. Further speculations in this direction have included, for instance, supersymmetric models~\cite{Yang:2022gvz,Du:2022pbp,Tang:2022pxh,Athron:2022isz,Zheng:2022irz,Ghoshal:2022vzo}, non-supersymmetric extended Higgs sectors~\cite{Lu:2022bgw,Fan:2022dck,Zhu:2022scj,Song:2022xts,Mondal:2022xdy,Ghosh:2022zqs,Bahl:2022xzi,Heo:2022dey,Babu:2022pdn,Biekotter:2022abc,Ahn:2022xeq,Han:2022juu,Arcadi:2022dmt,Lee:2022gyf,Ghorbani:2022vtv,Batra:2022pej,Batra:2022org,Popov:2022ldh}, vector-like fermions~\cite{Lee:2022nqz,Kim:2022zhj,Kawamura:2022uft,Crivellin:2022fdf,Nagao:2022oin}, leptoquarks~\cite{Bhaskar:2022vgk,Cheung:2022zsb,Athron:2022qpo} and SM effective field theory~\cite{Bagnaschi:2022whn,DiLuzio:2022xns,Cirigliano:2022qdm,Gupta:2022lrt}. 

In this work, we aim to offer an explanation of the two aforementioned anomalies using Type-II seesaw as the basic framework. 
Pedagogy dictates to look at the minimal Type-II framework first. And it is seen that despite its attractiveness, the Type-II seesaw model predicts a negative muon magnetic moment~\cite{Fukuyama:2009xk}, and hence, cannot account for the observed discrepancy. And this can be  understood from the chirality structure of the Yukawa interactions of the scalar triplet. This calls for extending the minimal Type-II model. And the extension used in this work comprises
a doubly charged $SU(2)_L$ singlet scalar andvector-like leptons (VLLs), first introduced in \cite{Chakrabarty:2020jro}. The VLLs can have novel origins such as Grand Unification~\cite{Thomas:1998wy,Freitas:2020ttd} and
the SM suitably
augmented by VLLs can in fact explain the muon $g-2$ anomaly~\cite{Kannike:2011ng,Dermisek:2013gta,Megias:2017dzd,Crivellin:2018qmi}. However, the minimal VLL scenario does not offer solutions to the neutrino mass and vacuum instability problems, Moreover, it gets rather constrained by the measurements of the Higgs to dimuon decay made by ATLAS~\cite{Aad:2020xfq} and CMS~\cite{Sirunyan:2018hbu}. Some other recent studies employing vector leptons and additional scalar multiplets to address the muon anomaly are~\cite{Frank:2020smf,Chun:2020uzw,Chen:2020tfr,Jana:2020joi,deJesus:2020upp,newzhou:2022cql}.

A doubly charged scalar is an ingredient of certain classes of BSM scenarii, the minimal left-right symmetric model (LRSM) augmented with scalar triplets being an example. That is, the triplets $\Delta_L$ (\textbf{1,3,1},2) and $\Delta_R$ (\textbf{1,1,3},2) are introduced under the LRSM gauge group $SU(3)_c \times SU(2)_L \times SU(2)_R \times U(1)_{B-L}$ \cite{Iso:2009nw}, over and above the minimal field content. On the other hand, some investigations involving a scalar triplet and VLLs are~\cite{Bahrami:2013bsa,Bahrami:2015mwa,Bahrami:2016has}. We thus have two doubly charged scalars in this scenario instead of one as in the case of ordinary Type-II seesaw\footnote{\cite{Chakrabarty:2018qtt} presents explanations the muon anomaly in models featuring a two doubly charged scalars but no additional fermions over and above the SM ones.}. The VLLs include both doublets and singlets under $SU(2)_L$, the latter carrying one unit of electric charge.
We show in this study how a positive contribution of the sought
magnitude to the muon $g-2$ can be obtained in this framework through a non-zero mixing of the two doubly charged bosons. We also demonstrate that tuning the Yukawa interactions and the triplet vacuum expectation value (VEV) correctly can help evade the constraints coming from
the non-observation of charged lepton flavour violation (CLFV)~\cite{Calibbi:2017uvl}. In addition, we compute the one-loop RG equations corresponding to this model and subsequently show that a stable EW vacuum exists within the parameter region  that reproduces the observed pattern of neutrino masses and mixings and accommodates the muon and CDF II anomalies.

This paper is organised as follows. We detail in field content and the corresponding interactions in section \ref{model}. The relevant constraints are listed in section \ref{constraints}. We review the chirality-flip in the muon $g-2$
amplitude in section \ref{gmt_and_MW} demonstrate how the observed $M_W$ can be simultaneously achieved. Section \ref{vacstab} presents an analysis combining vacuum stability, muon $g-2$ and the various relevant constraints. We summarise in section \ref{summary}. Various important formulae are relegated to the Appendix. 

\section{The model
}\label{model}
The minimal Type-II seesaw model employs a $SU(2)_L$ complex scalar triplet $\Delta$ in addition to the scalar doublet $\phi$. We invoke the field content introduced in \cite{Chakrabarty:2020jro} wherein the minimal Type-II model sector is augmented by a doubly charged scalar singlet $k^{++}$ and the following VLL multiplets:
\besub
\bea
L_{L,R} = 
\begin{pmatrix}
N_{L,R} \\
E_{L,R}
\end{pmatrix},~~~~
E^\prime_{L,R}.
\eea
\eesub
The quantum numbers of the BSM fields are shown in Table~\ref{bsm}. The scalar doublet and the triplet can be parameterised as under.
\besub
\bea
\phi = 
\begin{pmatrix}
    \phi^+ \\
    \frac{1}{\sqrt{2}}(v + \phi_0 + i \eta_0)
  \end{pmatrix},~~~~~
\Delta = 
\begin{pmatrix}
   \frac{\delta^+}{\sqrt{2}} & \delta^{++} \\
    \frac{1}{\sqrt{2}}(v_\D + \delta_0 + i \chi_0) & 
    -\frac{\delta^+}{\sqrt{2}}.
  \end{pmatrix}  
\eea
\eesub
Here, $v$ and $v_\Delta$ denote the VEVs acquired by the 
CP-even neutral components of $\phi$ and $\Delta$ respectively.

\begin{table}
\centering
\begin{tabular}{ |c|c| } 
\hline
Field & $SU(3)_c \times SU(2)_L \times U(1)_Y$ \\ 
\hline \hline 
$\Delta$ & $(1,3,1)$ \\ \hline 
$k^{++}$ & $(1,1,2)$\\ \hline 
$L_{L,R}$ & $(1,2,-1/2)$\\ \hline 
$E^\prime_{L,R}$ & $(\mathbf{1,1},-1)$\\ \hline 
\end{tabular}
\caption{Quantum numbers of the relevant fields under the SM gauge group.}
\label{bsm}
\end{table}

As for how the additional fields interact, we first show the scalar potential below:
\bea
V &=& V_2 + V_3 + V_4,
\eea
where $V_2,V_3$ and $V_4$ respectively contain the operators of mass dimension 2,3 and 4. That is,
\besub
\bea
V_2 &=& \m^2_{\phi} (\phi^{\dagger} \phi) + M^2_{\Delta} \text{Tr} (\Delta^{\dagger}\Delta) + M^2_S |k^{++}|^2, 
\\
V_3 &=& \mu_1^{}\, \phi^T (i \sigma_2^{}) \Delta^{\dagger} \phi 
 + \mu_2^{}\, \text{Tr}\big( \Delta^{\dagger} \Delta^{\dagger}\big) k^{++} + \text{h.c.} \\
V_4 &=& \frac{\l}{2} (\phi^{\dagger} \phi)^2
+ \frac{\l_1}{2} [\text{Tr} (\Delta^{\dagger}\Delta)]^2
+ \frac{\l_2}{2}\Big([\text{Tr} (\Delta^{\dagger}\Delta)]^2 - \text{Tr}(\Delta^{\dagger}\Delta\Delta^{\dagger}\Delta) \Big)  + \frac{\l_3}{2} |k^{++}|^4 \nonumber \\
&&
+ \l_4 \phi^{\dagger} \phi \text{Tr} 
(\Delta^{\dagger}\Delta)
+ \l_5 \phi^{\dagger} \big[\Delta, \Delta^\dagger \big] \phi + \l_6 \phi^{\dagger}\phi |k^{++}|^2
+ \l_7 \text{Tr}(\Delta^{\dagger}\Delta) |k^{++}|^2 \nonumber \\
&&
+ \l_8^{} \big({\tilde{\phi}}^{\dagger} \Delta \phi k^{--} +  \text{h.c.}\big). 
\eea 
\eesub
All parameters in the scalar potential are chosen real to rule out CP-violation. Electroweak symmetry breaking (EWSB) leads to mixings between the component scalars of 
$\Delta$ and $\phi$, as well as between $\Delta$ and $k^{++}$, subject to the conservation of electric charge and CP. The mass eigenstates thus emerging are the CP-even $(h,H)$, the CP-odd $A$, the singly charged $H^+$, and the doubly charged $(H_1^{++},H_2^{++})$. The constraint on the $\rho$-parameter dictates $v_\D \lesssim 4$ GeV in which case mixings involving the neutral and singly charged states can be rendered negligibly small. The corresponding physical masses in this limit are
\besub
\bea
M_h^2 &\simeq& \l v^2, \\
M_H^2 &=& M_A^2 \simeq M^2_\D + \frac{1}{2}(\l_4 + \l_5)v^2, \\
M^2_{H^+} &=&  M^2_\D + \frac{1}{2}\l_4 v^2.
\eea
\eesub
On the other hand, $\l_8 \neq 0$ can lead to a sizeable mixing between the doubly charged states despite the stringent upper limit on $v_\D$. The mass terms then have the following form for $v_\Delta << v$ and $\mu_2 << \l_8 v$:
\besub
\bea
\mathcal{L}^{++}_m &=& 
\begin{pmatrix}
\delta^{--} & k^{--}
\end{pmatrix}
\begin{pmatrix}
M^2_{\Delta} + \frac{1}{2} \l_4^{} v^2 & \frac{1}{2}\l_8 v^2 \\
\frac{1}{2}\l_8 v^2 & M^2_S
 + \frac{1}{2} \l_6 v^2
\end{pmatrix}
\begin{pmatrix}
\delta^{++} \\
 k^{++}
\end{pmatrix}\label{matrix_doublycharged}
\eea
\eesub 

We diagonalise the mass matrix in Eq.(\ref{matrix_doublycharged}) by rotating ($\delta^{++},k^{++}$) by an angle $\theta$ and obtain the mass eigenstates $H_{1,2}^{++}$ having masses 
$M^{++}_{1,2}$. Thus,
\besub
\bea
\begin{pmatrix}
    \delta^{++} \\
     k^{++}
  \end{pmatrix} =
\begin{pmatrix}
    \text{cos}\theta & \text{sin}\theta \\
    -\text{sin}\theta & \text{cos}\theta
  \end{pmatrix}
\begin{pmatrix}
    H_1^{++} \\
    H_2^{++}
  \end{pmatrix}
\eea
\eesub

We also give below the expressions for $M^{++}_{1,2}$ and $\theta$ in the $v_\Delta < < v$ limit:
\besub
\bea
(M_{1,2}^{++})^2 &=& \frac{1}{2}\big[(A + B) \pm 
\sqrt{(A - B)^2 + 4 C^2}\big]\label{mpp} ~, \\
\text{tan} 2\theta &=& \frac{2 C}{B - A} ~, ~~~\text{where}\label{tan} \\
A &=& M^2_{\Delta} + \frac{1}{2} \l_4 v^2 ~,\label{Aprime} \\
B &=& M^2_S + \frac{1}{2} \l_6^{} v^2 ~,\label{Bprime} \\
C &=& \frac{1}{2}\l_8^{} v^2 ~.\label{Cprime} 
\eea
\eesub

We discuss the fermionic interactions next. First, the bare mass terms of the VLLs and their interactions with the Higgs doublet $\phi$ are
\besub
\bea
\mathcal{L}^{\text{VLL}}_{Y,\phi} &=& -M \bar{L_L} L_R - M^\prime \bar{E_L^\prime} E_R^\prime - y_4 \bar{L}_L \phi E_R^\prime 
- y_4^\prime \bar{L}_R \phi E_L^\prime + \text{h.c.}
\eea
\eesub
The mixings of the VLLs with the SM leptons
are neglected in this work for simplicity\footnote{The mixings, even if allowed, are rendered small from the non-observation of CLFV. This has been explicitly demonstrated in \cite{Ishiwata:2013gma} for VLLs having quantum numbers identical to the present scenario. Therefore, they anyway do not majorly modify the muon $g-2$ prediction in this model thereby justifying the choice. Other constraints on such mixings, although subleading to CLFV, stem from the measurement of $p p \to h \to \mu \mu$ \cite{Dermisek:2013gta} and $p p \to h \to 4 l$ \cite{Dermisek:2014cia}.}. The mass terms of the VLLs then take the form
\besub
\bea
\mathcal{L}^{\text{VLL}}_{Y,\phi} \supset
-\begin{pmatrix}
 \bar{E_R} & \bar{E^\prime_R}
 \end{pmatrix}
\begin{pmatrix}
    M & \frac{y^\prime_4 v}{\sqrt{2}} \\
    \frac{y_4 v}{\sqrt{2}} & M^\prime
\end{pmatrix}  
\begin{pmatrix}
E_L \\
E^\prime_L
 \end{pmatrix} + \text{h.c.} \label{VLL_mat}
\eea
\eesub
The non-hermitian matrix in Eq.(\ref{VLL_mat}) is diagonalisable by the bi-unitary transformation
\bea
{U_R}^\dagger M_V U_L &=& M_V^d,
\eea
where 
\bea
M_V = 
\begin{pmatrix}
    M & \frac{y^\prime_4 v}{\sqrt{2}} \\
    \frac{y_4 v}{\sqrt{2}} & M^\prime
\end{pmatrix}, ~~~
M_V^d = 
\begin{pmatrix}
    M_1 & 0 \\
    0 & M_2
\end{pmatrix}~~~\text{and} ~~~
U_{L(R)} = 
\begin{pmatrix}
    \text{cos}\a_{L(R)} & \text{sin}\a_{L(R)} \\
    -\text{sin}\a_{L(R)} & \text{cos}\a_{L(R)}
  \end{pmatrix}.
\eea
Therefore, the VLLs in the mass basis, i.e.,  
$E_{L(R)_1}$ and $E_{L(R)_2}$, are obtained by rotating the flavour basis as 
\besub
\bea
\begin{pmatrix}
    E_{L(R)} \\
    E_{L(R)}^\prime
  \end{pmatrix} =
U_{L(R)}
\begin{pmatrix}
    E_{{L(R)}_1} \\
    E_{{L(R)}_2}
  \end{pmatrix}.
\eea
\eesub

Taken up next are the Yukawa interactions involving the triplet $\Delta$.
Denoting an SM lepton doublet (singlet) as $L_{\a L}$ ($l_{\a R}$), one writes
\besub
\bea
\mathcal{L}_{Y,\Delta} &=& \mathcal{L}^{\text{SM}}_{Y,\Delta} + \mathcal{L}^\text{VLL}_{Y,\Delta}, \\
\mathcal{L}^{\text{SM}}_{Y,\Delta} &=& -\sum_{\a,\b=e,\mu,\tau}
y_{\Delta}^{\a \b} \bar{L^c_{\a L}}~i\sigma_2 \Delta~L_{\b L} + \text{h.c.}, \\
\mathcal{L}^\text{VLL}_{Y,\Delta} &=& -2\sum_{\a =e,\mu,\tau}
y_{\Delta}^{\a 4} \bar{L^c_{\a L}}~i\sigma_2 \Delta~L_L 
-y_{\Delta}^{4 4} \bar{L^c_{L}}~i\sigma_2 \Delta~L_L
+ \text{h.c.}
\eea
\eesub
One notes that
$\mathcal{L}^{\text{VLL}}_{Y,\Delta}$ describes how the VLLs interact with $\Delta$
and is an addition over the minimal Type-II. Finally, we describe the Yukawa interactions involving $k^{++}$ below.
\besub
\bea
\mathcal{L}_{Y,k^{++}} &=& \mathcal{L}^{\text{SM}}_{Y,k^{++}} + \mathcal{L}^\text{VLL}_{Y,k^{++}}, \\
\mathcal{L}^{\text{SM}}_{Y,k^{++}} &=&
-\sum_{\a,\b=e,\mu,\tau}
y_{S}^{\a \b}~\bar{l^c_{\a R}}~l_{\b R} k^{++}
 + \text{h.c.}, \\
\mathcal{L}^{\text{VLL}}_{Y,k^{++}} &=&
-2\sum_{\a=e,\mu,\tau} 
y_S^{\a 4}~\bar{l^c_{\a R}}~E_R^\prime k^{++}
- y_S^{4 4}~\bar{{E^\prime}^c_{R}}~E_R^\prime k^{++}
 + \text{h.c.} 
\eea
\eesub

It is convenient to describe a framework in terms of masses and mixing angles. The following scalar quartic couplings are solved in terms of physical scalar masses and the mixing angle $\theta$ as under.
\besub
\bea
\l &=& \frac{M^2_h}{v^2}, \label{l}\\
\l_4 &=& \frac{2(M^2_{H^+} - M^2_\Delta)}{v^2}, \label{l4}\\
\l_5 &=& \frac{2(M^2_H - M^2_{H^+})}{v^2}, \label{l5}\\
\l_6 &=& \frac{2\big[ (M_1^{++})^2 \text{sin}^2\theta + 
(M_2^{++})^2 \text{cos}^2\theta - M_S^2 \big]}{v^2}, \label{l6}\\
\l_8 &=& \frac{2 \text{sin}\theta \text{cos}\theta
\big[ (M_2^{++})^2 - (M_1^{++})^2 \big]}{v^2}\label{l8}. 
\eea
\eesub
The independent parameters in the scalar sector are therefore $\{v,v_\Delta,\mu_2,M_h,M_H,M_H^+,M_1^{++},M_2^{++},M_\Delta,M_S,\l_1,\l_2,\l_3,\l_7\}$. Of these, we fix $M_h$ = 125 GeV and $v \simeq$ 246 GeV for $v_\Delta << v$.

A non-zero $v_\Delta$ leads to non-zero neutrino-mass elements of the form $m_\nu^{\a\b} = \sqrt{2} 
y_\Delta ^{\a\b} v_\Delta$. This necessitates $y_\Delta^{\a\b}$ to be complex. All other Yukawa couplings
are taken real since they do not participate in neutrino mass generation. One can also eliminate $y_4,y_4^\prime$ in favour of the VLL masses and $\a_L,\a_R$ as
\besub
\bea
y_4 &=& \frac{\sqrt2}{v}( M_2 \text{sin}\a_L~\text{cos}\a_R- M_1 \text{cos}\a_L~\text{sin}\a_R  ), \label{y4} \\
y_4^\prime &=& \frac{\sqrt2}{v}( M_2 \text{cos}\a_L~\text{sin}\a_R  - M_1\text{sin}\a_L~\text{cos}\a_R  ). \label{y4p}
\eea
\eesub
The neutral member of the VLL multiplet, $N$, then has
the mass
\bea
M_N = M = M_1 \text{cos} \a_L \text{cos} \a_R + M_2 \text{sin} \a_L \text{sin} \a_R.
\eea
The independent paramaters in the fermionic sector are therefore $\{m_\nu^{\a\b},y_\Delta^{\a 4},y_\Delta^{4 4},y_S^{\a \b},y_S^{\a 4},y_S^{4 4},M_1,M_2,\a_L,\a_R\}$ of which $m_\nu^{\a\b}$ are determined by the neutrino-oscillation data.

\section{constraints} \label{constraints}
We discuss here the constraints on the model from theory and experiments.

\subsection{Theoretical constraints}

The bounds $|\l_i| \leq 4\pi, |y_i| < \sqrt{4\pi}$ ensure that the theory remains perturbative, where $\l_i$ ($y_i$) denotes a generic quartic (Yukawa) coupling. In addition, the following conditions ensure a bounded-from-below (BFB) scalar potential for large field values of the constituent scalar fields:
\besub
\bea
\l > 0,~\l_1 > 0,~2\l_1 + \l_2 > 0,~\l_3 > 0, \label{v1}\\
\l_4 \pm \l_5 + \sqrt{\l \l_1} > 0,~\sqrt{\l \big(\l_1 + \frac{\l_2}{2} \big)} > 0,~\l_4 \pm \l_5 + 
\sqrt{\l \big(\l_1 + \frac{\l_2}{2} \big)} > 0, \label{v2}\\
\l_6 + \sqrt{\l \l_3} > 0,
~\l_7 + \sqrt{\l_1 \l_3} > 0,
~\l_7 + \sqrt{\l_3 \big(\l_1 + \frac{\l_2}{2} \big)} > 0\label{v3}.
\eea
\eesub
A given condition in the aforementioned set comes from demanding the 
scalar potential remains BFB in a given direction in the field space.
Unitarity leads to additional constraints on the quartic couplings. 
A tree-level 2 $\to$ 2
scattering matrix can be constructed between various two particle states consisting of charged
and neutral scalars~\cite{PhysRevD.7.3111,PhysRevD.16.1519}. Unitarity demands that the absolute value of each eigenvalue of
the aforementioned  matrix must be bounded from above at 
8$\pi$. The conditions for the present scenario are derived in \cite{Chakrabarty:2020jro} and are given below\footnote{The results have been checked with \cite{PhysRevD.84.095005} in the appropriate limit.}
\besub
\bea
|\l_1 + \l_2| \leq 8\pi,~|\l_4 - 2\l_5| \leq 8\pi,~|\l_4 \pm \l_5| \leq 8\pi, 
~|2 \l_1 + 3\l_2| \leq 16\pi, \label{u1}\\
\Big(\l + \l_1 - \l_2 \pm \sqrt{(\l - \l_1 + \l_2)^2 + 16 \l_5^2} \Big) \leq 16\pi, \label{u2}\\
\Big(\l + \l_7 \pm \sqrt{(\l - \l_7)^2 + 8 \l_8^2} \Big) 
\leq 16\pi, \label{u3}\\
\Big(\l_4 + 2\l_5 + \l_6 \pm \sqrt{(\l_4 - 2\l_5 - \l_6)^2 + 24 \l_8^2} \Big) \leq 16\pi\label{u4}.
\eea
\eesub
Finally, these bounds obtained from demanding perturbativity and
a BFB as well as unitary scalar potential are imposed at each energy scale in the analysis using RGEs. 

\subsection{Neutrino mass}
The $U_{\text{PMNS}}$ matrix diagonalizes the neutrino mass matrix $m_\nu$, {\it i.e.},
\besub
\bea
&&
m_\nu = U_{\text{PMNS}}^* ~m_\nu^{\text{diag}} ~U_{\text{PMNS}}^T
~, \label{nu}\\
&& \text{with}
~U_{\text{PMNS}} = V_{\text{PMNS}} \times 
\text{diag}(1,e^{i \a_{21}/2},,e^{i \a_{31}/2}) ~\mbox{and} \\
&& V_{\text{PMNS}} =  \begin{pmatrix}
    c_{12} c_{13} & s_{12} c_{13} &  s_{13} e^{-i \delta_{CP}}\\
 -s_{12}c_{23} - c_{12}s_{23}s_{13} e^{i \delta_{CP}} & c_{12}c_{23} - s_{12}s_{23}s_{13} e^{i \delta_{CP}} & s_{23}c_{13} \\
  s_{12}s_{23}-c_{12}c_{23}s_{13} e^{i \delta_{CP}} 
  & -c_{12}s_{23} - s_{12}c_{23}s_{13} e^{i \delta_{CP}} & c_{23} c_{13}
  \end{pmatrix}
  ~,
\eea
\eesub
where $s_{ij} =\sin\theta_{ij}$, $c_{ij} =\cos\theta_{ij}$, $\delta_{CP}^{}$ is the Dirac phase, and $\a_{21}^{}$ and $\a_{31}^{}$ are the Majorana phases. The neutrino oscillation parameters are fixed to their central values~\cite{Patrignani:2016xqp} as
\bea
&&
\text{sin}^2\theta_{12} = 0.307,~ 
~\text{sin}^2\theta_{23} = 0.510,~
~\text{sin}^2\theta_{13} = 0.021, \nonumber \\
&&
\Delta m^2_{21}
 = 7.45 \times 10^{-5} ~\text{GeV}^2,
~\Delta m^2_{32} = 2.53 \times 10^{-3} ~\text{GeV}^2,  \nonumber \\
&&
\delta_{CP} =  1.41\pi,~ 
~\a_{21} = \a_{31} = 0. \label{nuparam_fixed}
\eea
The mass of the lightest neutrino and Majorana phases are additionally taken zero in
the present analysis.

\subsection{Collider constraints}

Limits on the VLL masses are weak for negligible mixing of the VLLs with the SM leptons, which is the case here. A limit from colliders for a heavy charged lepton quotes $M,M^\prime > 102.6$ GeV~\cite{PhysRevD.98.030001}. A weak limit $\sim \mathcal{O}$(MeV) on masses neutral leptons comes from Big Bang Nucleosynthesis (BBN)~\cite{PhysRevD.98.030001}. We therefore take $M_N,M_1,M_2 > 110$ GeV in the present analysis. Next we discuss possible exclusion bounds on the doubly charged scalar masses. We find that stringent limits such as 
$M_{1,2}^{++} \gtrsim $ 1 TeV apply for doubly charged scalars dominantly decaying to an $l^+ l^+$ pair~\cite{ATLAS:2022yzd,CMS:2017pet}. The bounds are much weaker when the $W^+W^+$ mode dominates and this occurs for $v_\Delta > 10^{-4}$ GeV in the minimal Type-II model~\cite{ATLAS:2021jol}.

The NP sector in this study is further constrained by the Higgs signal strength data from the LHC. And the most stringent constraint for negligible $\phi_0-\delta_0$ mixing comes from the measurement of the $h \to \gamma\gamma$ signal strength. The presence of additional charged scalars and leptons implies that additional one-loop contributions to the $h\gamma\gamma$ amplitude shall arise thereby modifying  
the corresponding decay width \emph{w.r.t.} the SM. The amplitude stemming from the charged scalars $H^+,H_{1,2}^{++}$ and the VLLs $E^\pm_{1,2}$ is given by~\cite{Djouadi:2005gi,Djouadi:2005gj,Arhrib:2011vc}
\besub
\bea
\mathcal{M}^{\text{NP}}_{h \to \gamma \gamma} &=& \sum_{S = H^+, H_1^{++}, H_2^{++}}
q^2_S~\frac{\l_{h S S^*} v}{2 M^2_S} A_0\bigg(\frac{M^2_h}{4 M^2_S}\bigg) + \sum_{i=1,2} y_{h E_i E_i} A_{1/2}\Big(\frac{M^2_h}{4 M^2_i}\Big)
\label{htogaga_np}
\eea
\eesub
Where, in addition to $q_{H^+} = 1,~q_{H_{1,2}^{++}} = 2$,
\besub
\bea
\l_{h H^+ H^-} &=& \l_4 v, \\
\l_{h H_1^{++} H_1^{--}} &=& v \big\{(\l_4 - \l_5) 
c^2_\theta + \l_6 s^2_\theta - 2 \l_8 s_\theta c_\theta\big\}, \\
\l_{h H_1^{++} H_2^{--}} &=& v \big\{(\l_4 - \l_5) 
s^2_\theta + \l_6 c^2_\theta + 2 \l_8 s_\theta c_\theta\big\}, \\
y_{h E_1 E_1} &=& \frac{1}{2 v} \Big[ M_1 
\Big(-1 + \text{cos}(2\alpha_L) \text{cos}(2\alpha_R) \Big) + M_2 \Big( \text{sin}(2\alpha_L) \text{sin}(2\alpha_R) \Big)\Big], \\
y_{h E_2 E_2} &=& \frac{1}{2 v} \Big[ M_2 
\Big(-1 + \text{cos}(2\alpha_L) \text{cos}(2\alpha_R) \Big) + M_1 \Big( \text{sin}(2\alpha_L) \text{sin}(2\alpha_R) \Big)\Big]. 
\eea
\eesub

The total amplitude and the decay width then become
\bea
\mathcal{M}_{h \to \gamma \gamma} &=& 
\mathcal{M}^{\text{SM}}_{h \to \gamma \gamma} +
\mathcal{M}^{\text{NP}}_{h \to \gamma \gamma}, 
\\
\Gamma_{h \to \gamma \gamma} &=& \frac{G_F \a_{em}^2 M_h^3}{128 \sqrt{2} \pi^3} |\mathcal{M}_{h \to \gamma \gamma}|^2.\eea
where $G_F$ and $\a_{em}$ denote respectively the Fermi constant and the QED fine-structure constant. The various loop functions are expressed below~\cite{Djouadi:2005gj}.
\besub
\bea
A_{1/2}(x) &=& \frac{2}{x^2}\big((x + (x -1)f(x)\big), \\
A_0(x) &=& -\frac{1}{x^2}\big(x - f(x)\big),  \\
\text{with} ~~f(x) &=& \text{arcsin}^2(\sqrt{x}); ~~~x \leq 1 
\nonumber \\
&&
= -\frac{1}{4}\Bigg[\text{log}\frac{1+\sqrt{1 - x^{-1}}}{1-\sqrt{1 - x^{-1}}} -i\pi\Bigg]^2; ~~~x > 1.
\eea
\eesub
where $A_0(x)$ and $A_{1/2}(x)$ are the amplitudes for the spin-0 and spin-$\frac{1}{2}$ particles in the loop respectively.
The signal strength for the $\gamma\gamma$ channel is defined as
\bea
\mu_{\gamma\gamma} &=&  \frac{\sigma(pp \to h)\text{BR}(h \to \gamma \gamma)}{\Big[\sigma(pp \to h)\text{BR}(h \to \gamma \gamma)\Big]_{\text{SM}}}
\eea
Given the new scalars and VLLs do not modify the $pp\to h$ production rate,
\bea
\mu_{\gamma\gamma} &=&  \frac{\text{BR}(h \to \gamma \gamma)}{\Big[\text{BR}(h \to \gamma \gamma)\Big]_{\text{SM}}}, \\
& \simeq & \frac{\Gamma^{\text{SM}}_{h\to\gamma\gamma}}{\Gamma_{h\to\gamma\gamma}}.
\eea 
The latest 13 TeV results on the diphoton signal strength from
the LHC read $\mu_{\gamma\gamma} = 0.99^{+0.14}_{-0.14}$ (ATLAS~\cite{Aaboud:2018xdt}) and
$\mu_{\gamma\gamma} = 1.18^{+0.17}_{-0.14}$ (CMS~\cite{Sirunyan:2018ouh}).
Upon using the standard combination of signal strengths and uncertainties, we obtain $\mu_{\gamma\gamma} = 1.06 \pm 0.1$ and impose this constraint at 2$\sigma$. 

\subsection{Charged lepton flavour violation}

The presence of both singly and doubly charged scalars in this framework implies that charged lepton flavour violating (CLFV) processes of the form $l_\a \to l_\b \bar{l_\gamma} l_\delta$ and $l_\alpha \to l_\beta \gamma$ are triggerred at the tree- and one-loop level respectively. However, non observation of CLFV in various experiments has led to stringent upper bounds on the corresponding branching ratios. In this work, we abide by such bounds the latest of which are summarised in Table \ref{lfv_bound} below.
\begin{table}
\centering
\begin{tabular}{ |c|c| } 
\hline
 LFV channel & Experimental bound \\ 
 \hline \hline 
  $\mu \rightarrow e \gamma$ & $<$ 4.2 $\times 10^{-13}$ 
 ~\cite{TheMEG:2016wtm} \\ \hline
   $\tau \rightarrow e \gamma$ & $<$ 1.5 $\times 10^{-8}$ 
 ~\cite{Aubert:2009ag} \\ \hline
   $\tau \rightarrow \mu \gamma$ & $<$ 1.5 $\times 10^{-8}$ 
 ~\cite{Aubert:2009ag} \\ \hline
  $\mu \rightarrow \bar{e} e e$ & $<$ 1 $\times 10^{-12}$ 
 ~\cite{Bellgardt:1987du} \\ \hline
  $\tau \rightarrow \bar{e} e e$ & $<$ 1.4 $\times 10^{-8}$ 
 ~\cite{Amhis:2016xyh} \\ \hline
 $\tau \rightarrow \bar{\mu} e e$ & $<$ 8.4 $\times 10^{-9}$ 
 ~\cite{Amhis:2016xyh} \\ \hline
  $\tau \rightarrow \bar{\mu} e \mu$ & $<$ 1.6 $\times 10^{-8}$ 
 ~\cite{Amhis:2016xyh} \\ \hline
  $\tau \rightarrow \bar{e} \mu \mu$ & $<$ 9.8 $\times 10^{-9}$ 
 ~\cite{Amhis:2016xyh} \\ \hline
   $\tau \rightarrow \bar{e} \mu e$ & $<$ 1.1 $\times 10^{-8}$ 
 ~\cite{Amhis:2016xyh} \\ \hline
   $\tau \rightarrow \bar{\mu} \mu \mu$ & $<$ 1.2 $\times 10^{-8}$ 
 ~\cite{Amhis:2016xyh} \\ \hline
\end{tabular}
\caption{Latest upper limits on LFV branching ratios.}
\label{lfv_bound}
\end{table}

\section{The $W$-mass, muon $g-2$ amplitude and charged lepton flavour violation}\label{gmt_and_MW}

The $W$-mass predicted by this model can be expressed in terms of the NP contributions to the oblique parameters $\Delta S$, $\Delta T$ and $\Delta U$ as~\cite{Maksymyk:1993zm}
\bea
M^2_W &=& M^2_{W,\text{SM}} \bigg[1 + \frac{\alpha_{em}}{c^2_W - s^2_W} 
\bigg( -\frac{\Delta S}{2} + c^2_W \Delta T
+ \frac{c^2_W - s^2_W}{4 s^2_W} \Delta U \bigg)  \bigg]
\eea
where $c_W$ and $\a_{em}$ respectively denote the cosine of the Weinberg angle and the
fine-structure constant. A given oblique parameter $X$ in this framework receives NP contributions from scalars $\Delta$ and $k^{++}$ sector as well as the VLLs. That is, $\Delta X = \Delta X_{\Delta,k^{++}} + \Delta X_{\text{VLL}}$. One expresses below $\Delta S_{\text{VLL}}$ and $\Delta T_{\text{VLL}}$ following \cite{Anastasiou:2009rv,PhysRevD.47.2046,PhysRevD.70.015003,Chen:2017hak}:
\besub
\bea
\Delta S_{\text{VLL}} &=& \frac{1}{2\pi} \Bigg[ \sum_i \Big(
 (|A^L_{N i}|^2 + |A^R_{N i}|^2) \Psi_+(x_N,x_i) + 2 A^L_{N i} A^R_{N j} \Psi_-(x_N,x_i) \Big) \nonumber \\
&&
+ \sum_{ij} \Big( - \frac{1}{2}(|X^L_{i j}|^2 + |X^R_{i j}|^2) \chi_+(x_i,x_j)
+ X^L_{ij} X^R_{ij} \chi_-(x_i,x_j) \Big) \Bigg], \\
\Delta T_{\text{VLL}} &=& \frac{1}{16\pi s^2_W c^2_W} \Bigg[ \sum_i \Big(
 (|A^L_{N i}|^2 + |A^R_{N i}|^2) \theta_+(x_N,x_i) + 2 A^L_{N i} A^R_{N j} \theta_-(x_N,x_i) \Big) \nonumber \\
&&
+ \sum_{ij} \Big( - \frac{1}{2}(|X^L_{i j}|^2 + |X^R_{i j}|^2) \theta_+(x_i,x_j)
+ X^L_{ij} X^R_{ij} \theta_-(x_i,x_j) \Big) \Bigg]
\eea
\eesub
where $x_N = \frac{M^2_N}{M^2_Z}$ and $x_{1,2} = \frac{M^2_{1,2}}{M^2_Z}$.
The forms of the various functions are given in the Appendix. One further derives for the present scenario:
\besub
\bea
A^{L(R)}_{N 1} = c_{L(R)},~A^{L(R)}_{N 2} = s_{L(R)}, \\
X^{L(R)}_{11} = -c^2_{L(R)},~X^{L(R)}_{12} = X^{L(R)}_{21} = -c_{L(R)}s_{L(R)},~X^{L(R)}_{22} = -s^2_{L(R)}.
\eea
\eesub
Next, we come to discussing the oblique corrections coming from the scalar sector. We remind that a ($t_3,Y$)\footnote{We use the $Q=t_3 + Y$ convention.} scalar couples with $Z$ and $\gamma$ with the coefficients $e(t_3 c^2_W - Y s^2_W)/(c_W s_W)$ and $e(t_3 + Y)$ respectively~\cite{Lavoura:1993nq}. The two doubly charged scalars 
$\delta^{++}$ and $k^{++}$ respectively carry 
($t_3^{(1)}=1,Y^{(1)}=1$) and ($t_3^{(2)}=0,Y^{(2)}=2$). The interations connecting $H_{1,2}^{++}$ to $Z$ and $\gamma$ can be parameterised as
\bea
\mathcal{L^{\text{doubly charged}}_{\text{gauge}}} &=& i e \Big( (\partial_\mu H^{++}_i) H^{--}_j - (\partial_\mu H^{--}_i) H^{++}_j \Big)(z_{ij} Z^\mu + a_{ij} A^\mu) \nonumber \\
&&
 + e^2 H^{++}_i H^{--}_j
(m_{ij} Z_\mu Z^\mu + n_{ij} Z_\mu A^\mu + p_{ij} A_\mu A^\mu).
\eea
We find $p_{ij} = 2a_{ij} = 4 \delta_{ij}$. The other coefficients are expressed as $2\times2$ matrices below.
\bea
[z] &=& \begin{pmatrix}
(-1 + 2 c_{2W} + c_{2\t})/(s_{2W}) &&  s_{2\t}/ s_{2W} \\
s_{2\t}/ s_{2W} && (1 + 2 c_{2W} - c_{2\t})/(s_{2W})
\end{pmatrix},
\eea


\bea
[m] &=& \begin{pmatrix}
    (c^2_{2W}/s^4_W c^2_\t + 4 s^2_\t) t^2_W & c_\t s_\t (1/s^2_W - 3/c^2_W) \\
   c_\t s_\t (1/s^2_W - 3/c^2_W)  & (c^2_{2W}/s^4_W s^2_\t + 4 c^2_\t) t^2_W
  \end{pmatrix},
\eea  

\bea
[n] &=& \begin{pmatrix}
    2(-1 + 2 c_{2W} + c_{2\t})/(s_W c_W) & 2 s_{2\t}/(s_W c_W) \\
   2 s_{2\t}/(s_W c_W)  & -2(1-2 c_{2W} + c_{2\t})/(c_W s_W)
  \end{pmatrix}.
\eea
We express below the gauge-boson self-energies coming from scalar loops in terms of the Passarino-Veltman (P-V) functions~\cite{PASSARINO1979151}.
\besub
\bea
\Pi^{\Delta,k^{++}}_{ZZ}(p^2) &=& -4 e^2 \sum_{i,j=1,2} 
z^2_{ij} B_{00}\big(p^2,(M^{++}_i)^2,
(M^{++}_j)^2\big) - 4 e^2 \frac{s^2_W}{c^2_W} B_{00}\big(p^2,M^2_{H^+},
M^2_{H^+}\big) \nonumber \\
&& 
 - 4 e^2 \frac{1}{s^2_W c^2_W} B_{00}\big(p^2,M^2_H,
M^2_H\big) + 2 e^2 \sum_{i=1,2} m_{ii} A_0\big( (M^{++}_i)^2 \big) \nonumber
\\
&&
 + 2 e^2 \frac{s^2_W}{c^2_W } A_0\big( M^2_{H^+} \big)
 + 2 e^2 \frac{1}{c^2_W s^2_W} A_0\big( M^2_H \big), \\ \nonumber \\
 \Pi^{\Delta,k^{++}}_{\gamma\gamma}(p^2) &=& -4 e^2 \sum_{i,j=1,2} 
a^2_{ij} B_{00}\big(p^2,(M^{++}_i)^2,
(M^{++}_j)^2\big) - 4 e^2 B_{00}\big(p^2,M^2_{H^+},
M^2_{H^+}\big) \nonumber \\
&& 
+ 2 e^2 \sum_{i=1,2} p_{ii} A_0\big( (M^{++}_i)^2 \big) 
 + 2 e^2 A_0\big( M^2_{H^+} \big),\\ \nonumber \\
  \Pi^{\Delta,k^{++}}_{\gamma Z}(p^2) &=& -4 e^2 \sum_{i,j=1,2} 
a_{ij} z_{ij} B_{00}\big(p^2,(M^{++}_i)^2,
(M^{++}_j)^2\big) + 4 e^2 \frac{s_W}{c_W} B_{00}\big(p^2,M^2_{H^+},
M^2_{H^+}\big)  \nonumber \\
&&
+ e^2 \sum_{i=1,2} n_{ii} A_0\big( (M^{++}_i)^2 \big) 
 - 2 e^2 \frac{s_W}{c_W} A_0\big( M^2_{H^+} \big).
 \eea
\eesub
The P-V functions are evaluated using the publicly available library \texttt{LoopTools}~\cite{HAHN1999153}. The scalar sector contribution to the 
$S$-parameter reads
\bea
\frac{\a_{em}}{4 c^2_W s^2_W}\Delta S_{\Delta,k^{++}} = \frac{\Pi^{\Delta,k^{++}}_{ZZ}(M^2_Z)-\Pi^{\Delta,k^{++}}_{ZZ}(0)}{M^2_Z}
 - \frac{\partial \Pi^{\Delta,k^{++}}_{\gamma\gamma}(p^2)}{\partial p^2}\bigg|_{p^2=0}
+ \frac{c^2_W - s^2_W}{c_W s_W} \frac{\partial \Pi^{\Delta,k^{++}}_{\gamma Z}(p^2)}{\partial p^2}\bigg|_{p^2=0}. 
\eea

Now $\Delta T_{\Delta,k^{++}}$ has a 
counterterm at quantum level induced by the counter term of $v_\Delta$, say $\delta v_\Delta$. And $\delta v_\Delta$ is not unique and depends on the choice of the renormalisation scheme instead. For example, $\delta v_\Delta$ can always be adjusted to cancel potentially large contributions from the scalar loops. Therefore, we set $\Delta T_{\Delta,k^{++}} = 0$ in this work.

For $\Delta U = 0$, global fit studies of the EW parameters \cite{Asadi:2022xiy,Lu:2022bgw} in light of the CDF II anomaly reveal the following allowed ranges and correlation for $\Delta S$ and $\Delta T$:
\bea
\Delta S = 0.15 \pm 0.08,~~\Delta T = 0.27 \pm 0.06,~~\rho_{ST} = 0.93.
\eea

Next, we discuss the muon anomalous magnetic moment for this setup.
It is reminded that while $\delta^{++}$ and $k^{++}$ respectively couple to only left chiral and right chiral leptons, the mass eigenstates $H_1^{++}$ and $H_2^{++}$ couple to both chiralities. That is, the interactions of muons with the VLLs and 
$H^{++}_{1,2}$ can be expressed as
\bea
\mathcal{L} &=& 2\sum_{i=1,2} \sum_{j=1,2} \bar{\mu}^c
(y_L^{ij} P_L + y_R^{ij} P_R) E_i H_j^{++} + \text{h.c.},
\eea
where
\bea
y_L^{11} = y_\Delta^{\mu 4} \text{cos}\a_L \text{cos}\theta,
~y_R^{11} = -y_S^{\mu 4} \text{sin}\a_R \text{sin}\theta,
~y_L^{12} = y_\Delta^{\mu 4} \text{cos}\a_L \text{sin}\theta, 
~y_R^{12} = y_S^{\mu 4} \text{sin}\a_R \text{cos}\theta, \nonumber \\
y_L^{21} = y_\Delta^{\mu 4} \text{sin}\a_L \text{cos}\theta, 
~y_R^{21} = y_S^{\mu 4} \text{cos}\a_R 
\text{sin}\theta, 
~y_L^{22} = y_\Delta^{\mu 4} \text{sin}\a_L \text{sin}\theta, 
~y_R^{22} = -y_S^{\mu 4} \text{cos}\a_R 
\text{cos}\theta.
\eea

We assume $y_\Delta^{e 4},y_\Delta^{\tau 4},
y_\Delta^{44},y_S^{\a\b},y_S^{e 4},y_S^{\tau 4},
y_S^{44}$ to be vanishingly small\footnote{Demanding $\Delta,k^{++}$ and the VLLs to be odd under some 
$\mathbb{Z}_2$ symmetry while keeping the SM fields even under the same necessitates 
$y_\Delta^{44},y_S^{\a\b},
y_S^{44} = 0.$ We refer to the last paragraph of section \ref{model} for a discussion.}. 
The one-loop muon $g-2$ $\Delta a_\mu$ has the 
following three distinct components in this limit:
\bea
\Delta a_\mu = \big(\Delta a_\mu^{+}\big)
_{l}
 + \big(\Delta a_\mu^{++}\big)_{l}
 + \big(\Delta a_\mu^{++}\big)_{\text{VLL}}.\label{delta_amu}
\eea

In Eq.(\ref{delta_amu}), 
$\big(\Delta a_\mu^{+}\big)_{l}$ 
($\big(\Delta a_\mu^{++}\big)_{l}$) denotes the contribution from the one-loop amplitude involving SM leptons + singly (doubly) charged scalars. The expression for 
$\big(\Delta a_\mu^{+}\big)
_{l}$ is given by~\cite{Fukuyama:2009xk}
\besub
\bea
\big(\Delta a_\mu^{+}\big)
_{l} &=& - \frac{m^2_\mu}{8\pi^2}
\frac{v^2}{v^2 + 2 v^2_\Delta}
\sum_{\a = e,\mu,\tau} \big(y_\Delta^\dagger U_{\text{PMNS}}^* \big)_{\mu \a} 
\big( U^T_{\text{PMNS}} y_\Delta \big)_{\a\mu}\nonumber \\
&&
\int_0^1 dx \frac{x^2(1-x)}{m^2_\mu x^2 + (M^2_{H^+} - m^2_\mu
-m^2_\a)x + m^2_\a} \label{g-2_II_singly}, \\
& \simeq & -\frac{\big(m^2_\nu\big)_{\mu\mu}}{96\pi^2}
\frac{m^2_\mu}{v_\Delta^2 M^2_{H^+}}.
\eea
\eesub
The contribution from $H^+$ is thus negative and identical to the minimal Type-II seesaw.
Also,
\besub
\bea
\big(\Delta a_\mu^{++}\big)_{l} &=&
-\frac{m_\mu^2}{8\pi^2} \sum_{i} \sum_{\a} b_i  
\big( y^\dagger_\Delta \big)_{\mu\a} \big(y_\Delta\big)_{\a\mu} \nonumber \\
&&
\int_0^1 dx \Bigg[ \frac{4x^2(1-x)}{m^2_\mu x^2 + 
((M^{++}_i)^2 - m^2_\mu
-m^2_\a)x + m^2_\a} \nonumber \\
&&
+ \frac{2x^2(1-x)}{m^2_\mu x^2 + (m^2_\a - m^2_\mu
-(M^{++}_i)^2)x + (M^{++}_i)^2} \Bigg], \label{g-2_II_doubly} \\
\text{where} ~b_1 = c^2_\theta,~b_2 = s^2_\theta
\nonumber \\
& \simeq & -\frac{\big(m^2_\nu\big)_{\mu\mu}}{12\pi^2}
\frac{m^2_\mu}{v_\Delta^2} \Bigg( \frac{c^2_\theta}{(M_1^{++})^2} + \frac{s^2_\theta}{(M_2^{++})^2} \Bigg).
\eea
\eesub
The contribution involving the SM leptons and the doubly charged scalars is also found negative. The completely left-chiral couplings between the charged scalars and the SM leptons entails no chirality-flip must occur thereby predicting a negative contribution to the $g-2$ amplitude.

\begin{figure}[tbhp]
\begin{center}
\includegraphics[scale=0.50]{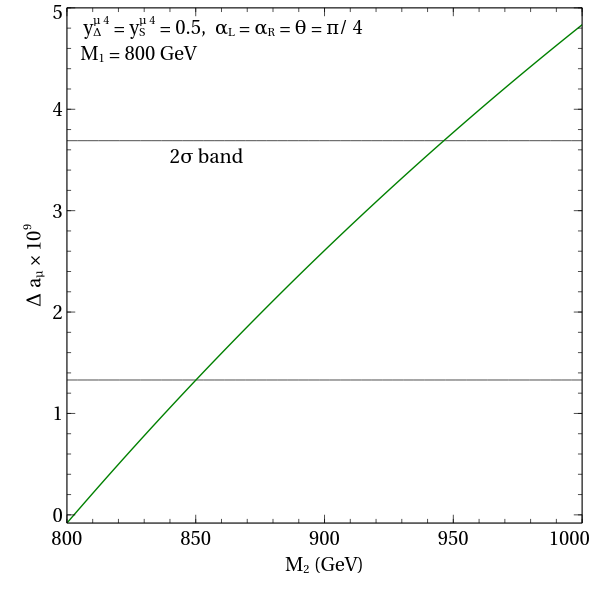}~~~~
\includegraphics[scale=0.50]{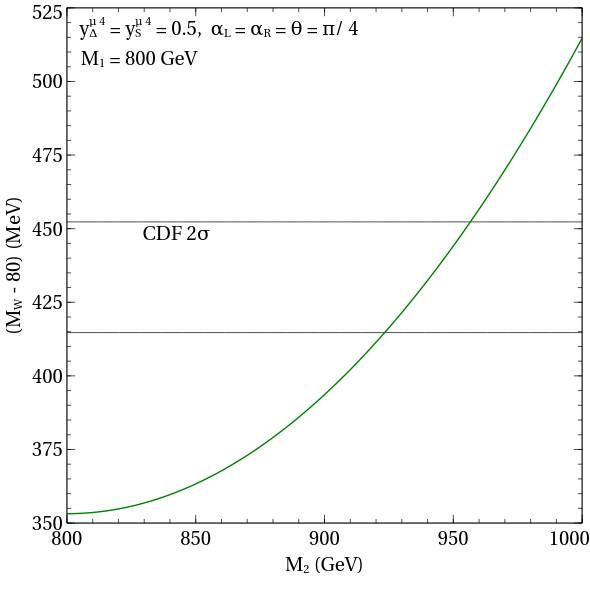}
\caption{The variation of $\Delta a_\mu$ and $M_W$ with $M_2$ for $M_1^{++}$ = 500 GeV and $\Delta M^{++}$ = 50 GeV. The values for the other parameters are given in the panels.}
\label{f:g-2}
\end{center}
\end{figure}

The contribution from the VLLs is\footnote{An excellent review containing analytical formulae for $\Delta a_\mu$  for generic classes of models is \cite{Lindner:2016bgg}}
\bea
\big(\Delta a_\mu^{++}\big)_{\text{VLL}} &=&
\sum_{i=1,2} \sum_{j=1,2} \Bigg[ -\frac{m^2_\mu}{4\pi^2} \Big(\{(y_L^{ij})^2 + (y_L^{ij})^2\} I_1(M_i,M_j^{++}) + \frac{2 M_i}{m_\mu} y_L^{ij} y_R^{ij} I_2(M_i,M_j^{++}) \Big) \nonumber \\
&&
-\frac{m^2_\mu}{2\pi^2} \Big(\{(y_L^{ij})^2 + (y_R^{ij})^2\} I_3(M_i,M_j^{++}) + \frac{2 M_i}{m_\mu} y_L^{ij} y_R^{ij}
I_4(M_i,M_j^{++}) \Big) \Bigg].
\label{g-2_VLL}
\eea
The integrals $I_{a}(m_1,m_2),~a=1,2,3,4$ for $m_\mu << M_{1,2},~M^{++}_{1,2}$ are
\besub
\bea
I_1(m_1,m_2) &=& \int_0^1 dx~\frac{x^2(1-x)}{m_1^2 x + m_2^2(1-x)}, \\
I_2(m_1,m_2) &=& \int_0^1 dx~\frac{x^2}{m_1^2 x + m_2^2(1-x)}, \\
I_3(m_1,m_2) &=& \int_0^1 dx~\frac{x^2(1-x)}{m_1^2 (1-x) + m_2^2 x}, \\
I_4(m_1,m_2) &=& \int_0^1 dx~\frac{x(1-x)}{m_1^2 (1-x) + m_2^2 x}.
\eea
\eesub
Given the integrals $I_{a}(m_1,m_2),~a=1,2,3,4$ are all positive, the contribution to $\big(\Delta a_\mu^{++}\big)_{\text{VLL}}$
from the first and third terms in Eq.(\ref{g-2_VLL}) are negative.  
On the other hand, a chirality flip is identified in the second and fourth terms. In fact, the chirality-flipping component of 
$\big(\Delta a_\mu^{++}\big)_{\text{VLL}}$ can be extracted upon 
defining $\Delta M \equiv M_2 - M_1 << M_1$ and $\Delta M^{++} \equiv M_2^{++} - M_1^{++} << M_1^{++}$. We additionally take $\a_L = \a_R$ for simplicity. The chirality-flipping component in its lowest order of $\Delta M$ and $\Delta M^{++}$ then becomes
\besub
\bea
\big(\Delta a_\mu^{++}\big)_{\text{VLL}}^{\text{cf}} &\simeq& \frac{m_\mu}{4\pi^2}y_{\Delta}^{\mu 4}y_{S}^{\mu 4} \text{sin}(2\a_R)\text{sin}(2\theta) \frac{\Delta M \Delta M^{++}}{(M_1^{++})^3} f\Big( \frac{M_1^2}{(M_1^{++})^2} \Big), \label{cf} \\
f(r) &=& \frac{(-35 r^3 + 15 r^2 + 27 r - 7) + (12 r^3 + 40 r^2 - 2 r - 1)\text{log}(r)}{2(r-1)^5}.
\eea
\eesub
Two important observations that emerge are that (i) the chirality flipping component is enhanced \emph{w.r.t.} the chirality preserving part of $\big(\Delta a_\mu^{++}\big)_{\text{VLL}}$ and the Type-II like terms by an $\mathcal{O} \Big(\frac{M_i}{m_\mu} \Big)$ factor, and (ii) the chirality flip can be of either sign. It is therefore possible to generate a positive contribution to $\Delta a_\mu$ of the observed magnitude by choosing the relevant parameters appropriately.
We also inspect that the magnitude of $\big(\Delta a_\mu^{++}\big)_{\text{VLL}}^{\text{cf}}$ is maximised for $\theta = \frac{\pi}{4}$ and $\a_L = \frac{\pi}{4}$ when $\a_L = \a_R$. We shall adhere to these values in this section while quantifying $M_W$ and $\Delta a_\mu$. While showing the variation of $\Delta a_\mu$ and $M_W$ \emph{w.r.t.} $M_2$ in Fig.\ref{f:g-2}, we additionally fix $M_1^{++}$ = 500 GeV, 
$\Delta M^{++}$ = 50 GeV, $\big(y_\Delta^{\mu 4},y_S^{\mu 4}\big)$ = (0.5,0.5) and $v_\D = 10^{-3}$ GeV. We mention here that the choice of $v_\Delta$ and the particle masses is consistent with the LHC excusion bounds and signal strength constraints.
The following are the takeaways from FIG.\ref{f:g-2}. 
Increasing $M_2$ while holding $M_1$ and the other parameters fixed accordingly increases $\Delta a_\mu$ and potentially puts it in the sought range. For $M_1$ = 800 GeV, it is seen that $\Delta a_\mu$ is in the 2$\sigma$ range for 850 GeV $\lesssim M_2 \lesssim$ 940 GeV. On the other hand, the larger the mass splitting between the VLLs $E_1$ and $E_2$, the larger the corresponding $T$-parameter 
$\Delta T_{\text{VLL}}$ and hence, the larger is the predicted value of $M_W$. This explains the monotonically increasing curve on the right panel. In fact, $M_W$ lies in the 2$\sigma$ corresponding to the CDF II observation for 920 GeV $\lesssim M_2 \lesssim$ 955 GeV. For the said choice of the other parameters,
one thus identifies 920 GeV $\lesssim M_2 \lesssim$ 940 GeV as the band that simultaneously resolves both the muon $g-2$ and CDF II anomalies in this model. It is important to remind here that while $M_W$ is not sensitive to $y_\Delta^{\mu 4}$ and $y_S^{\mu 4}$, $\big(\Delta a_\mu^{++}\big)_{\text{VLL}}^{\text{cf}}$ simply scales with the product of two Yukawas. We chose here $\big(y_\Delta^{\mu 4},y_S^{\mu 4}\big)$ = (0.5,0.5) as a reference and the behaviour for other values can be predicted easily.
\begin{figure}[tbhp]
\begin{center}
\includegraphics[scale=0.50]{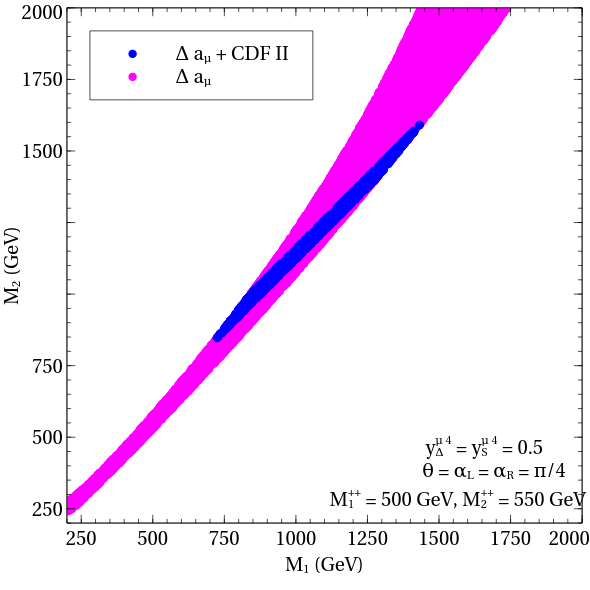}~~~~
\includegraphics[scale=0.50]{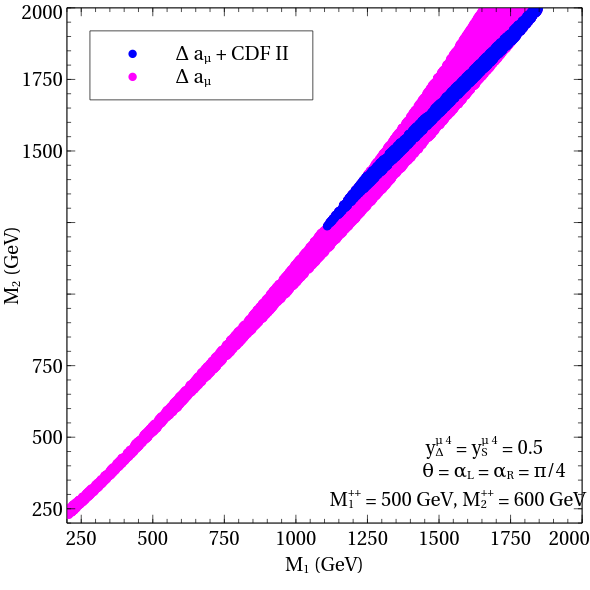}~~~~ 
\\
\includegraphics[scale=0.50]{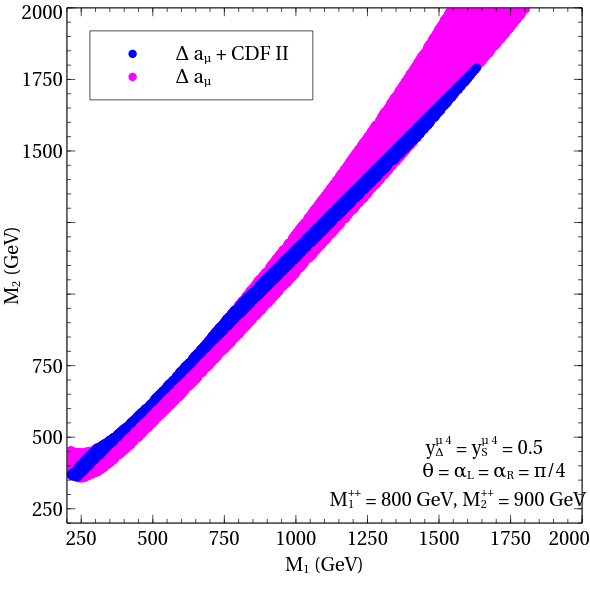}~~~~
\caption{Parameter regions in the $M_1-M_2$ compliant with all the constraints as well as the 2$\sigma$ ranges of muon $g-2$ (magenta) and $M_W$ (blue). The values taken by the other parameters are shown in the plots.}
\label{f:M1_M2}
\end{center}
\end{figure}
We show in FIG.~\ref{f:M1_M2} the parameter regions in the $M_1-M_2$ plane compatible with the two anomalies for specific choices for the other relevant parameters (as given in the plots). One reads from eqn.(\ref{cf}) that $y_\Delta^{\mu 4}$ and $y_S^{\mu 4}$, $\big(\Delta a_\mu^{++}\big)_{\text{VLL}}^{\text{cf}}$ $\propto\Delta 
M \Delta M^{++}$. Increasing $\Delta M$ from 
50 GeV to 100 GeV keeping the $y_\Delta^{\mu 4},y_S^{\mu 4}, M_1^{++}$ fixed would accordingly demand a larger $\Delta M$ so as to maintain $\big(\Delta a_\mu^{++}\big)_{\text{VLL}}^{\text{cf}}$ at the same value. This is why the band compatible with the observed $\Delta a_\mu$ in the $M_1-M_2$ plane gets thinner while moving from the top left to the top right panel of FIG.~\ref{f:M1_M2}. On the other hand, while $M_1^{++}$ is increased from 500 GeV to 800 GeV with $\Delta M^{++}$ fixed at 100 GeV, the chirality-flip contribution tends to decrease thereby entailing an increase in 
$\Delta M$ so as to predict the same $\Delta a_\mu$ as before.

The assumption that $y_\Delta^{e 4},y_\Delta^{\tau 4}$, 
$y_S^{\b 4}$ are suppressed\footnote{Even if no such approximation is a priori made, an estimation of the LFV chirality-flipping amplitude using \cite{Lindner:2016bgg} leads to $|y_{\Delta}^{\a 4}|, |y_S^{\a 4}| \sim \mathcal{O}(10^{-4})$ for $\a = e,\tau$ here.} forbids a similar $\mathcal{O}(M_i/m_\mu)$
chirality-flipped enhancement in $l_\a \to l_\b \gamma$. Analogously to $\big(\Delta a_\mu^{+}\big)_{l}$ and $\big(\Delta a_\mu^{++}\big)_{l}$,
a non-zero 
$l_a \to l_\b \gamma$ amplitude
is therefore induced only by the triplet 
$\Delta$. And since $\Delta$ only couples to the left-chiral leptons, the corresponding amolitude is qualitatively similar to as in the minimal Type-II model. The only difference comes from the fact that here we have two doubly charged scalars as opposed to the one in Type-II.
One then finds the corresponding branching ratio in the present model to be 
\bea
\text{BR}(l_\a \to l_\b \gamma) &=& \frac{\a_{em} 
|(m_\nu^2)_{\a\b}|^2}
{12\pi G_F^2 v_\Delta^4} \Bigg(\frac{1}{8 M^2_{H^+}}
 + \frac{c^2_\theta}{(M_1^{++})^2}
 + \frac{s^2_\theta}{(M_2^{++})^2} \Bigg)^2 \text{BR}(l_\a \to l_\b \nu_\a \bar{\nu_\b})\label{ltolgamma}.
\eea

Also, the branching ratios of the 3-body CLFV decays are given by \footnote{The corresponding formula for the Higgs triplet model is seen in \cite{Akeroyd:2009nu,Akeroyd:2009hb}}
\besub
\bea
\text{BR}(\mu \to \bar{e} e e) &=& \frac{|m_\nu^{e\mu}|^2 |m_\nu^{ee}|^2}{16 G^2_F v^4_\Delta} \Bigg(\frac{c^2_\theta}{(M_1^{++})^2}
 + \frac{s^2_\theta}{(M_2^{++})^2} \Bigg)^2 
\text{BR}(\mu \to e \bar{\nu_e} \nu_\mu), \\
\text{BR}(\tau \to \bar{l_{\a}} l_{\b} l_{\gamma}) &=& S\frac{|m_\nu^{\tau \a}|^2 |m_\nu^{\b \gamma}|^2}{16 G^2_F v^4_\Delta} \Bigg(\frac{c^2_\theta}{(M_1^{++})^2}
 + \frac{s^2_\theta}{(M_2^{++})^2} \Bigg)^2 
\text{BR}(\tau \to \mu \bar{\nu_\mu} \nu_\tau).
\eea
\eesub
In the above, $S$ = 1(2) for $\beta=\gamma$ ($\beta \neq \gamma$),
$\text{BR}(\mu \to e \bar{\nu_e} \nu_\mu)$ = 100$\%$ and $\text{BR}(\tau \to \mu \bar{\nu_\mu} \nu_\tau)$ = 17$\%$. 
We reckon that the branching ratios for all 
$\a,\b=e,\mu,\tau$ is controlled by $v_\Delta$ for fixed scalar masses. It is easy to check that the choice $v_\Delta = 10^{-3}$ GeV predicts branching ratios that are well below the CLFV bounds.

\section{Electroweak vacuum stability} \label{vacstab}

We search here for a stable EW vacuum till the Planck scale that is compliant with observed $\Delta a_\mu$ and $M_W$. The scalar potential along the direction of the scalar doublet is approximated as $V(\phi) = \frac{1}{4}\l(\phi) \phi^4$ for 
$\phi >> v$. All running couplings are to be evaluated at a scale 
$\mu = \phi$. The condition $\l(\mu) > 0$ ensures a stable EW vacuum. We choose $M_t$ = 172.76 GeV as the input scale from which the couplings start evolving and demand vacuum stability
up to the Planck scale. A complete set of the one-loop beta functions is given in the Appendix. One notes that $y_\Delta^{e 4}=y_\Delta^{\tau 4}=
y_\Delta^{44}=y_S^{\a\b}=y_S^{e 4}=y_S^{\tau 4}$=
$y_S^{44}$ = 0 is a fixed point of this model. It is therefore justified to neglect the effect of these parameters in the RGE since 
assigning tiny values to these at the EW scale ensures that they remain tiny at all scales. We now come to choosing the input scale values for the relevant couplings. First, for a given set of masses and mixing angles, the EW scale values for 
$\l,\l_4,\l_5,\l_6,\l_8$ and $y_4,y_4^\prime$ get fixed from Eqs.(\ref{l})-(\ref{l8}) and Eqs.(\ref{y4})-(\ref{y4p}) respectively. For the rest, we take $\l_1 = \l_2 = \l_7 = 0.01, \l_3 = 0.3$. We further fix $y^{\mu 4}_{\Delta}=y^{\mu 4}_{S}=0.5$ at the input scale, in compliance with the previous section. We interate here that we intend to be demonstrative of a stable vacuum in simultaneity with the observed $\Delta a_\mu$ and $M_W$, and therefore, refrain from performing an exhaustive scan of the model parameter space. For example, the EW scale value of 0.5 ensures that $y^{\mu 4}_{\Delta}=y^{\mu 4}_{S}$ gently increase with $\mu$ and remain well below the perturbative limit at the Planck scale. And this behaviour is not sensitive to the choice of the input scale quartic couplings since $\beta_{{y^{\mu 4}_\Delta}}$ and $\beta_{{y^{\mu 4}_S}}$ at one-loop are independent of the same. We equate $M_\Delta = M_S = M_H^+$ to $M_1^{++}$, the latter being 
a common mass scale here. As for the SM $t$-Yukawa and gauge couplings, we use $M_W$ = 80.384 and $\a_s(M_Z)$ = 0.1184 in which case the 
$t$-Yukawa and the gauge couplings at $\mu = M_t$ become 
$y_t = 0.93690,~g_1(\mu=M_t) = 0.3583,
~g_2 = 0.6478,
~g_3 = 1.1666$~\cite{Buttazzo:2013uya}.
The requirements of a perturbative theory and a stable EW vacuum till the Planck scale are henceforth dubbed as 'high-scale validity'. Similar to the previous section, we take $v_\Delta = 10^{-3}$ GeV and run the following scan
\besub
\bea
200~\text{GeV} < M_1, M_2 < 2~\text{TeV}, \\
200~\text{GeV} < M^{++}_1, M^{++}_2 < 2~\text{TeV}, \\
0 < \theta < \frac{\pi}{2},
~0 < \a_L = \a_R < \frac{\pi}{2}.
\eea
\eesub
The lower limit of $M_{1,2}^{++}$ > 200 GeV is consistent with the LHC exclusion bounds for the $v_\Delta$ taken.
We look for parameter points that now predict a stable vacuum till $M_{\text{Pl}} = 1.22 \times 10^{19}$ GeV, in addition to accounting for $\Delta a_\mu$ and $M_W$ in their respective 2$\sigma$ bands. We remind that criteria of BFB-ness and unitarity defined by Eqs.(\ref{v1})-(\ref{v3}) and Eqs.(\ref{u1})-(\ref{u4}) respectively are also to be met at each intermediate scale in course of the RG evolution.
\begin{figure}[tbhp]
\begin{center}
\includegraphics[scale=0.50]{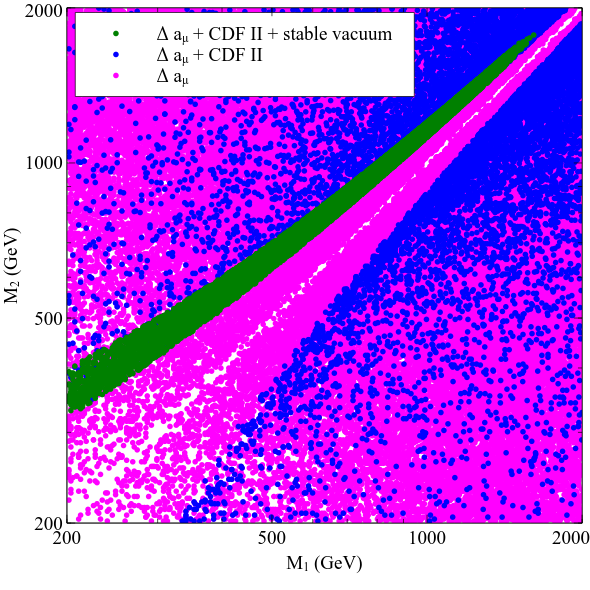}~~~~
\includegraphics[scale=0.50]{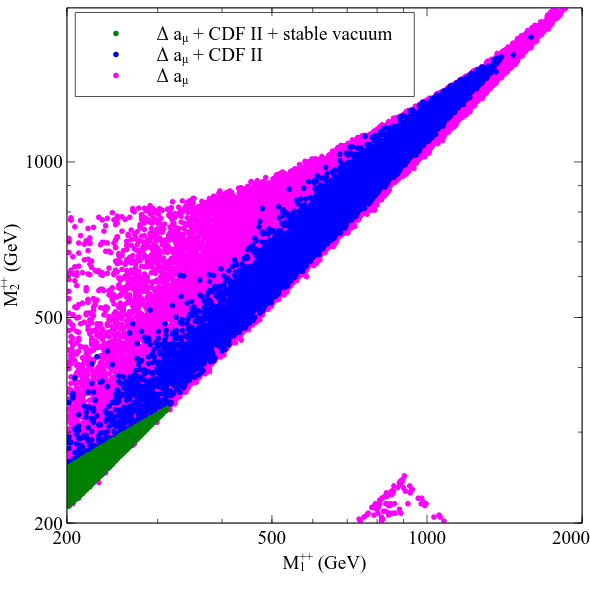}~~~~ \\
\includegraphics[scale=0.50]{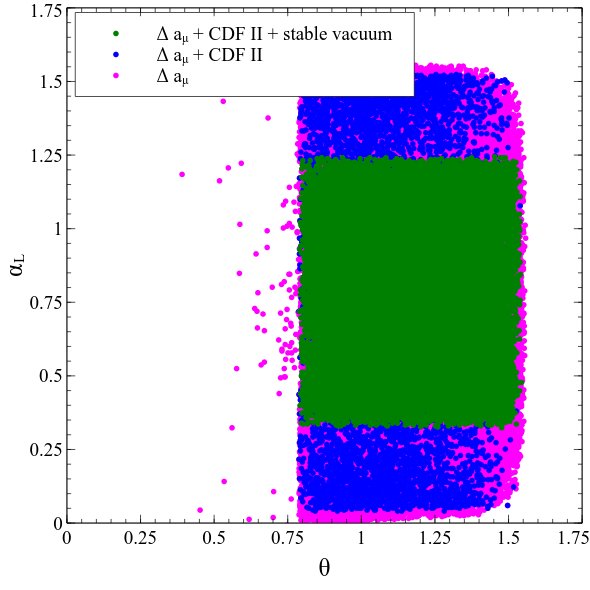}~~~~
\caption{Parameter points allowed by the various criteria in the $M_1-M_2$ (top left), $M^{++}_1-M^{++}_2$ (top right) and $\theta-\a_L$ (bottom) planes. The colour-coding is explained in the legends.}
\label{f:ano_vs}
\end{center}
\end{figure}
In Fig.\ref{f:ano_vs}, we present the parameter regions by the aforementioned criteria. We discover that a subset of parameter points predicting the sought values of $\Delta a_\mu$ and $M_W$
is also consistent of a stable vacuum till the Planck scale. We discuss the salient features thereby identified in the following. 
\begin{itemize}

\item Seeking an explanation of the muon $g-2$ anomaly does not majorly constrain the $M_1-M_2$ plane. Only a narrow strip around the $M_1=M_2$ line is excluded on grounds of requiring $\Delta M \neq 0$
which is necessary to obtain a 
$\Delta a_\mu$ in the required ballpark, as explained before. However, demanding an $M_W$ in the sought range widens the disallowed band around the $M_1=M_2$ line. This can be traced back to the oblique parameter values of the VLL sector. 

The most stringent constraint is derived from demanding a perturbative theory and a stable vacuum up to $M_{\text{Pl}}$. This can be understood as follows. Since the Yukawa couplings $y_4,y_4^\prime$ 
enter the $\beta$-functions of the theory, their EW scale values get tightly constrained. For instance, the magnitudes of these Yukawas get bounded from above from the $\l > 0$ condition since large Yukawas tend to destabilise the vacuum. Such bounds on $y_4,y_4^\prime$ translate to the observed constraint in the $M_1-M_2$ plane. In fact, the $0.35 \lesssim \a_L \lesssim 1.2$ bound also has the same origin.

\item The muon and CDF anomalies do not put bounds on the scalar masses themselves and instead constrain the mass-splitting at most. However, high-scale validity restricts $M_1^{++},M_2^{++} \lesssim 320$ GeV. These bounds can be traced back to the bounds on the boundary scale values of the quartic couplings $\l_6$ and $\l_8$ that get tightly 
restricted from high-scale validity. We note that there is no separate restriction on 
$\theta$ coming from high-scale validity.

\end{itemize}

\section{Conclusions}\label{summary}

The recently reported discrepancy in the measured value of $M_W$ has garnered significant attention in the particle physics community. In addition, the longstanding muon $g-2$ anomaly that was first reported by BNL, has also aroused a fresh interest in NP constructs after being corroborated by FNAL. In this work, we have proposed an explanation to both these anomalies in a common framework by extending the minimal Type-II seesaw model by by a doubly charged scalar singlet, an $SU(2)_L$ doublet of vector-like
leptons and a charged $SU(2)_L$ singlet vector-like lepton. We have demonstrated that an appropriate mass splitting between the charged VLLs can predict the observed muon anomalous magnetic moment through a chirality-flip. Simultaneouly, by virtue of the $S$- and $T$-parameters, the framework also leads to an $M_W$ consistent with the recent observation. In addition, we have also shown that the parameter region accommodating the two anomalies is consistent with the correct pattern of neutrino masses and mixings, lepton flavour violation and a stable EW vacuum up to the Planck scale.

\acknowledgments
NC acknowledges financial support from DST, India, under grant number IFA19-PH237 (INSPIRE Faculty Award).

\section{Appendix}

\subsection{Oblique parameter functions}
\besub
\bea
\psi_+(x,y) &=& \frac{1}{3} - \frac{1}{9}\text{log}\Big(\frac{x}{y} \Big), \\
\psi_-(x,y) &=& -\frac{x + y}{6\sqrt{x y}}, \\
\chi_+(x,y) &=& \frac{5(x^2 + y^2)-22 x y}{9(x-y)^2} + \frac{3 x y (x + y)-x^3-y^3}{3(x-y)^3}\text{log}\Big(\frac{x}{y} \Big), \\
\chi_-(x,y) &=& -\sqrt{x y} \bigg[ \frac{x + y}{6 x y} - \frac{x + y}{(x-y)^2} + \frac{2 x y}{(x-y)^2}\text{log}\Big(\frac{x}{y} \Big)
\bigg], \\
\theta_+(x,y) &=& x + y
 - \frac{2 x y}{x + y}\text{log}\Big(\frac{x}{y} \Big)\\
\theta_-(x,y) &=& 2\sqrt{x y} \Big[ \frac{x + y}{x - y}\text{log}\Big(\frac{x}{y} \Big) - 2 \Big].
\eea
\eesub

\subsection{One-loop beta functions}
The one-loop beta function for a quartic coupling $\l_i$ is split into scalar, gauge and fermionic terms as
$\beta_{\l_i} = \beta_{\l_i}^S + \beta_{\l_i}^g + \beta_{\l_i}^F$. Thus,
\besub
\bea
16 \pi^2 \beta^S_{\l} &=& 12 \l^2 + 6 \l_4^2 + 4 \l_5^2
+ 2 \l_6^2 + 4 \l_8^2, \\ 
16 \pi^2 \beta^S_{\l_1} &=& 14 \l_1^2 + 4 \l_1 \l_2 + 2 \l_2^2 + 4 \l_4^2 + 4 \l_5^2 + 2 \l_7^2, \\
16 \pi^2 \beta^S_{\l_2} &=& 12 \l_1 \l_2 + 3 \l_2^2 - 8 \l_5^2, \\
16 \pi^2 \beta^S_{\l_3} &=& 10 \l_3^2 + 4 \l_6^2 + 6 \l_7^2, \\
16 \pi^2 \beta^S_{\l_4} &=& 6 \l \l_4 + 8 \l_1 \l_4 
+ 2 \l_2 \l_4 + 4 \l_4^2 + 8 \l_5^2 + 2 \l_6 \l_7 + 4 \l_8^2, \\
16 \pi^2 \beta^S_{\l_5} &=& 2 \l \l_5 + 2 \l_1 \l_5
- 2\l_2 \l_5 + 8 \l_4 \l_5 - 4 \l_8^2, \\
16 \pi^2 \beta^S_{\l_6} &=& 6 \l \l_6 + 4 \l_3 \l_6
+ 4 \l_6^2 + 6 \l_4 \l_7 + 12 \l_8^2, \\
16 \pi^2 \beta^S_{\l_7} &=& 4 \l_4 \l_6 + 8 \l_1 \l_7
+ 2 \l_2 \l_7 + 4 \l_3 \l_7 + 4 \l_7^2 + 4 \l_8^2, \\
16 \pi^2 \beta^S_{\l_8} &=& 2 \l \l_8 + 4 \l_4 \l_8
- 8 \l_5 \l_8 + 4 \l_6 \l_8 + 2 \l_7 \l_8.
\eea
\eesub

\besub
\bea
16 \pi^2 \beta^g_{\l} &=& - 3\l(g_1^2 + 3 g_2^2) 
+ \frac{3}{4}g_1^4 + \frac{3}{4}g_1^2 g_2^2 
+ \frac{9}{4}g_2^4, \\
16 \pi^2 \beta^g_{\l_1} &=& - 12\l_1(g_1^2 + 2 g_2^2) 
+ 12 g_1^4 + 24 g_1^2 g_2^2 + 18 g_2^4,  \\
16 \pi^2 \beta^g_{\l_2} &=& - 12\l_2(g_1^2 + 2 g_2^2) 
- 48 g_1^2 g_2^2 
+ 12 g_2^4, \\
16 \pi^2 \beta^g_{\l_3} &=& - 48\l_3 g_1^2 + 192 g_1^4, \\
16 \pi^2 \beta^g_{\l_4} &=& - \l_4 
\Big( \frac{15}{2} g_1^2 + \frac{33}{2} g_2^2 \Big)
+ 3 g_1^4 + 6 g_2^4, \\
16 \pi^2 \beta^g_{\l_5} &=& - \l_5 
\Big( \frac{15}{2} g_1^2 + \frac{33}{2} g_2^2 \Big)
- 6 g_1^2 g_2^2, \\
16 \pi^2 \beta^g_{\l_6} &=& - \l_6\Big(\frac{51}{2} g_1^2 + 9 g_2^2 \Big) + 12 g_1^4, \\
16 \pi^2 \beta^g_{\l_7} &=& - \l_7\Big(20 g_1^2 + 8 g_2^2 \Big) + 48 g_1^4, \\
16 \pi^2 \beta^g_{\l_7} &=& - \l_8\Big(11 g_1^2 + 7 g_2^2 \Big) + 48 g_1^4.
\eea
\eesub

\besub
\bea
16 \pi^2 \beta^F_{\l} &=& 4\l \Big(3 y_t^2 + 3 y_b^2 +  y_\tau^2 +  y_4^2
+ (y_4^\prime)^2 \Big)
- 4 \Big(3 y_t^4 + 3 y_b^4 + y_\tau^4 + y_4^4
+ (y_4^\prime)^4 \Big), \\
16 \pi^2 \beta^F_{\l_1} &=& 16 \l_1 
\Big(y_\Delta^{\mu 4} \Big)^2 - 64 \Big(y_\Delta^{\mu 4} \Big)^4, \\
16 \pi^2 \beta^F_{\l_2} &=& 16 \l_2 
\Big(y_\Delta^{\mu 4} \Big)^2 + 64 \Big(y_\Delta^{\mu 4} \Big)^4, \\
16 \pi^2 \beta^F_{\l_3} &=& 16 \l_3 
\Big(y_S^{\mu 4} \Big)^2 - 64 \Big(y_S^{\mu 4} \Big)^4, \\
16 \pi^2 \beta^F_{\l_4} &=& \l_4 
\bigg[ \Big(8 y_\Delta^{\mu 4} \Big)^2 + 6 y_t^2
+ 6 y_b^2 + 2 y^2_\tau + 2 y_4^2 + 2 (y_4^\prime)^2 \bigg], \\
16 \pi^2 \beta^F_{\l_5} &=& \l_5 
\bigg[ \Big(8 y_\Delta^{\mu 4} \Big)^2 + 6 y_t^2
+ 6 y_b^2 + 2 y^2_\tau + 2 y_4^2 + 2 (y_4^\prime)^2 \bigg], \\
16 \pi^2 \beta^F_{\l_6} &=& \l_6 
\bigg[ \Big(8 y_\Delta^{\mu 4} \Big)^2 + 6 y_t^2
+ 6 y_b^2 + 2 y^2_\tau + 2 y_4^2 + 2 (y_4^\prime)^2 \bigg], \\
16 \pi^2 \beta^F_{\l_7} &=& \l_7 
\bigg[\Big(8 y_\Delta^{\mu 4} \Big)^2 + \Big(8 y_S^{\mu 4} \Big)^2 + 16 y_4^2 \Big(y_\Delta^{\mu 4} \Big)^2  \bigg], \\
16 \pi^2 \beta^F_{\l_8} &=& \l_8 
\bigg[\Big(4 y_\Delta^{\mu 4} \Big)^2 + \Big(4 y_S^{\mu 4} \Big)^2 + 6 y^2_t + 6 y^2_b + 2 y^2_\tau + 2 y_4^2 
+ (y_4^\prime)^2 \bigg].
\eea
\eesub

We next list the $\beta$-functions for the relevant Yukawa couplings below.
\besub
\bea
16 \pi^2 \beta_{y_t} &=& \frac{9}{2}y^3_t + y_t
\big(3 y_b^2 + y_\tau^2 + y^2_4 + (y^\prime_4)^2 - \frac{17}{12} g_1^2 - \frac{9}{4} g_2^2 - 8 g^2_3 \big),
\\
16 \pi^2 \beta_{y_b} &=& \frac{9}{2}y^3_b + y_b
\big(3 y_t^2 + y_\tau^2 + y^2_4 + (y^\prime_4)^2 - \frac{5}{12} g_1^2 - \frac{9}{4} g_2^2 - 8 g^2_3 \big), \\
16 \pi^2 \beta_{y_\tau} &=& \frac{5}{2}y^3_\tau
+ y_\tau
\big(3 y_t^2 + 3 y_b^2 + y^2_4 + (y^\prime_4)^2 - \frac{15}{4} g_1^2 - \frac{9}{4} g_2^2 \big), \\
16 \pi^2 \beta_{y_4} &=& \frac{5}{2}y^3_4
+ y_4
\big(3 y_t^2 + 3 y_b^2 + (y^\prime_4)^2
 - \frac{15}{4} g_1^2 - \frac{9}{4} g_2^2 \big), \\
16 \pi^2 \beta_{y_4^\prime} &=& \frac{5}{2}
(y_4^\prime)^3
+ y_4^\prime
\big(3 y_t^2 + 3 y_b^2 + y^2_4
 - \frac{15}{4} g_1^2 - \frac{9}{4} g_2^2 \big), \\
16 \pi^2 \beta_{{y^{\mu 4}_\Delta}} &=& 
8 (y_\Delta^{\mu 4})^3 + y_\Delta^{\mu 4} 
\big( \frac{y_4^2}{2} - \frac{3}{2} g_1^2
- \frac{9}{2} g_1^2 \big), \\
16 \pi^2 \beta_{{y^{\mu 4}_S}} &=& 
8 (y_S^{\mu 4})^3 + y_S^{\mu 4} 
\big( y_4^2 - 6 g_1^2 \big).
\eea
\eesub

Finally, the $\beta$-functions for the gauge couplings read
\besub
\bea
16 \pi^2 \beta_{g_1} &=& \frac{67}{6} g_1^3, \\
16 \pi^2 \beta_{g_2} &=& -\frac{13}{6} g_2^3, \\
16 \pi^2 \beta_{g_3} &=& -7 g_3^3.
\eea
\eesub

\bibliography{ref} 

\begin{thebibliography}{114}%
\makeatletter
\providecommand \@ifxundefined [1]{%
 \@ifx{#1\undefined}
}%
\providecommand \@ifnum [1]{%
 \ifnum #1\expandafter \@firstoftwo
 \else \expandafter \@secondoftwo
 \fi
}%
\providecommand \@ifx [1]{%
 \ifx #1\expandafter \@firstoftwo
 \else \expandafter \@secondoftwo
 \fi
}%
\providecommand \natexlab [1]{#1}%
\providecommand \enquote  [1]{``#1''}%
\providecommand \bibnamefont  [1]{#1}%
\providecommand \bibfnamefont [1]{#1}%
\providecommand \citenamefont [1]{#1}%
\providecommand \href@noop [0]{\@secondoftwo}%
\providecommand \href [0]{\begingroup \@sanitize@url \@href}%
\providecommand \@href[1]{\@@startlink{#1}\@@href}%
\providecommand \@@href[1]{\endgroup#1\@@endlink}%
\providecommand \@sanitize@url [0]{\catcode `\\12\catcode `\$12\catcode
  `\&12\catcode `\#12\catcode `\^12\catcode `\_12\catcode `\%12\relax}%
\providecommand \@@startlink[1]{}%
\providecommand \@@endlink[0]{}%
\providecommand \url  [0]{\begingroup\@sanitize@url \@url }%
\providecommand \@url [1]{\endgroup\@href {#1}{\urlprefix }}%
\providecommand \urlprefix  [0]{URL }%
\providecommand \Eprint [0]{\href }%
\providecommand \doibase [0]{http://dx.doi.org/}%
\providecommand \selectlanguage [0]{\@gobble}%
\providecommand \bibinfo  [0]{\@secondoftwo}%
\providecommand \bibfield  [0]{\@secondoftwo}%
\providecommand \translation [1]{[#1]}%
\providecommand \BibitemOpen [0]{}%
\providecommand \bibitemStop [0]{}%
\providecommand \bibitemNoStop [0]{.\EOS\space}%
\providecommand \EOS [0]{\spacefactor3000\relax}%
\providecommand \BibitemShut  [1]{\csname bibitem#1\endcsname}%
\let\auto@bib@innerbib\@empty
\bibitem [{\citenamefont {Chatrchyan}\ \emph {et~al.}(2012)\citenamefont
  {Chatrchyan} \emph {et~al.}}]{Chatrchyan:2012xdj}%
  \BibitemOpen
  \bibfield  {author} {\bibinfo {author} {\bibfnamefont {S.}~\bibnamefont
  {Chatrchyan}} \emph {et~al.} (\bibinfo {collaboration} {CMS}),\ }\href
  {\doibase 10.1016/j.physletb.2012.08.021} {\bibfield  {journal} {\bibinfo
  {journal} {Phys. Lett.}\ }\textbf {\bibinfo {volume} {B716}},\ \bibinfo
  {pages} {30} (\bibinfo {year} {2012})},\ \Eprint
  {http://arxiv.org/abs/1207.7235} {arXiv:1207.7235 [hep-ex]} \BibitemShut
  {NoStop}%
\bibitem [{\citenamefont {Aad}\ \emph {et~al.}(2012)\citenamefont {Aad} \emph
  {et~al.}}]{Aad:2012tfa}%
  \BibitemOpen
  \bibfield  {author} {\bibinfo {author} {\bibfnamefont {G.}~\bibnamefont
  {Aad}} \emph {et~al.} (\bibinfo {collaboration} {ATLAS}),\ }\href {\doibase
  10.1016/j.physletb.2012.08.020} {\bibfield  {journal} {\bibinfo  {journal}
  {Phys. Lett.}\ }\textbf {\bibinfo {volume} {B716}},\ \bibinfo {pages} {1}
  (\bibinfo {year} {2012})},\ \Eprint {http://arxiv.org/abs/1207.7214}
  {arXiv:1207.7214 [hep-ex]} \BibitemShut {NoStop}%
\bibitem [{\citenamefont {Degrassi}\ \emph {et~al.}(2012)\citenamefont
  {Degrassi}, \citenamefont {Di~Vita}, \citenamefont {Elias-Miro},
  \citenamefont {Espinosa}, \citenamefont {Giudice}, \citenamefont {Isidori},\
  and\ \citenamefont {Strumia}}]{Degrassi:2012ry}%
  \BibitemOpen
  \bibfield  {author} {\bibinfo {author} {\bibfnamefont {G.}~\bibnamefont
  {Degrassi}}, \bibinfo {author} {\bibfnamefont {S.}~\bibnamefont {Di~Vita}},
  \bibinfo {author} {\bibfnamefont {J.}~\bibnamefont {Elias-Miro}}, \bibinfo
  {author} {\bibfnamefont {J.~R.}\ \bibnamefont {Espinosa}}, \bibinfo {author}
  {\bibfnamefont {G.~F.}\ \bibnamefont {Giudice}}, \bibinfo {author}
  {\bibfnamefont {G.}~\bibnamefont {Isidori}}, \ and\ \bibinfo {author}
  {\bibfnamefont {A.}~\bibnamefont {Strumia}},\ }\href {\doibase
  10.1007/JHEP08(2012)098} {\bibfield  {journal} {\bibinfo  {journal} {JHEP}\
  }\textbf {\bibinfo {volume} {08}},\ \bibinfo {pages} {098} (\bibinfo {year}
  {2012})},\ \Eprint {http://arxiv.org/abs/1205.6497} {arXiv:1205.6497
  [hep-ph]} \BibitemShut {NoStop}%
\bibitem [{\citenamefont {Buttazzo}\ \emph {et~al.}(2013)\citenamefont
  {Buttazzo}, \citenamefont {Degrassi}, \citenamefont {Giardino}, \citenamefont
  {Giudice}, \citenamefont {Sala}, \citenamefont {Salvio},\ and\ \citenamefont
  {Strumia}}]{Buttazzo:2013uya}%
  \BibitemOpen
  \bibfield  {author} {\bibinfo {author} {\bibfnamefont {D.}~\bibnamefont
  {Buttazzo}}, \bibinfo {author} {\bibfnamefont {G.}~\bibnamefont {Degrassi}},
  \bibinfo {author} {\bibfnamefont {P.~P.}\ \bibnamefont {Giardino}}, \bibinfo
  {author} {\bibfnamefont {G.~F.}\ \bibnamefont {Giudice}}, \bibinfo {author}
  {\bibfnamefont {F.}~\bibnamefont {Sala}}, \bibinfo {author} {\bibfnamefont
  {A.}~\bibnamefont {Salvio}}, \ and\ \bibinfo {author} {\bibfnamefont
  {A.}~\bibnamefont {Strumia}},\ }\href {\doibase 10.1007/JHEP12(2013)089}
  {\bibfield  {journal} {\bibinfo  {journal} {JHEP}\ }\textbf {\bibinfo
  {volume} {12}},\ \bibinfo {pages} {089} (\bibinfo {year} {2013})},\ \Eprint
  {http://arxiv.org/abs/1307.3536} {arXiv:1307.3536 [hep-ph]} \BibitemShut
  {NoStop}%
\bibitem [{\citenamefont {Zoller}(2014)}]{Zoller:2014cka}%
  \BibitemOpen
  \bibfield  {author} {\bibinfo {author} {\bibfnamefont {M.~F.}\ \bibnamefont
  {Zoller}},\ }in\ \href@noop {} {\emph {\bibinfo {booktitle} {{17th
  International Moscow School of Physics and 42nd ITEP Winter School of Physics
  Moscow, Russia, February 11-18, 2014}}}}\ (\bibinfo {year} {2014})\ \Eprint
  {http://arxiv.org/abs/1411.2843} {arXiv:1411.2843 [hep-ph]} \BibitemShut
  {NoStop}%
\bibitem [{\citenamefont {Elias-Miro}\ \emph {et~al.}(2012)\citenamefont
  {Elias-Miro}, \citenamefont {Espinosa}, \citenamefont {Giudice},
  \citenamefont {Isidori}, \citenamefont {Riotto},\ and\ \citenamefont
  {Strumia}}]{EliasMiro:2011aa}%
  \BibitemOpen
  \bibfield  {author} {\bibinfo {author} {\bibfnamefont {J.}~\bibnamefont
  {Elias-Miro}}, \bibinfo {author} {\bibfnamefont {J.~R.}\ \bibnamefont
  {Espinosa}}, \bibinfo {author} {\bibfnamefont {G.~F.}\ \bibnamefont
  {Giudice}}, \bibinfo {author} {\bibfnamefont {G.}~\bibnamefont {Isidori}},
  \bibinfo {author} {\bibfnamefont {A.}~\bibnamefont {Riotto}}, \ and\ \bibinfo
  {author} {\bibfnamefont {A.}~\bibnamefont {Strumia}},\ }\href {\doibase
  10.1016/j.physletb.2012.02.013} {\bibfield  {journal} {\bibinfo  {journal}
  {Phys. Lett.}\ }\textbf {\bibinfo {volume} {B709}},\ \bibinfo {pages} {222}
  (\bibinfo {year} {2012})},\ \Eprint {http://arxiv.org/abs/1112.3022}
  {arXiv:1112.3022 [hep-ph]} \BibitemShut {NoStop}%
\bibitem [{\citenamefont {Isidori}\ \emph {et~al.}(2001)\citenamefont
  {Isidori}, \citenamefont {Ridolfi},\ and\ \citenamefont
  {Strumia}}]{Isidori:2001bm}%
  \BibitemOpen
  \bibfield  {author} {\bibinfo {author} {\bibfnamefont {G.}~\bibnamefont
  {Isidori}}, \bibinfo {author} {\bibfnamefont {G.}~\bibnamefont {Ridolfi}}, \
  and\ \bibinfo {author} {\bibfnamefont {A.}~\bibnamefont {Strumia}},\ }\href
  {\doibase 10.1016/S0550-3213(01)00302-9} {\bibfield  {journal} {\bibinfo
  {journal} {Nucl. Phys.}\ }\textbf {\bibinfo {volume} {B609}},\ \bibinfo
  {pages} {387} (\bibinfo {year} {2001})},\ \Eprint
  {http://arxiv.org/abs/hep-ph/0104016} {arXiv:hep-ph/0104016 [hep-ph]}
  \BibitemShut {NoStop}%
\bibitem [{\citenamefont {Swiezewska}(2016)}]{Swiezewska:2016rrp}%
  \BibitemOpen
  \bibfield  {author} {\bibinfo {author} {\bibfnamefont {B.~N.}\ \bibnamefont
  {Swiezewska}},\ }\emph {\bibinfo {title} {{Higgs boson and vacuum stability
  in models with extended scalar sector}}},\ \href
  {https://www.fuw.edu.pl/praca-doktorska/event5492.html} {Ph.D. thesis},\
  \bibinfo  {school} {Warsaw U.} (\bibinfo {year} {2016})\BibitemShut {NoStop}%
\bibitem [{\citenamefont {Schechter}\ and\ \citenamefont
  {Valle}(1980)}]{PhysRevD.22.2227}%
  \BibitemOpen
  \bibfield  {author} {\bibinfo {author} {\bibfnamefont {J.}~\bibnamefont
  {Schechter}}\ and\ \bibinfo {author} {\bibfnamefont {J.~W.~F.}\ \bibnamefont
  {Valle}},\ }\href {\doibase 10.1103/PhysRevD.22.2227} {\bibfield  {journal}
  {\bibinfo  {journal} {Phys. Rev. D}\ }\textbf {\bibinfo {volume} {22}},\
  \bibinfo {pages} {2227} (\bibinfo {year} {1980})}\BibitemShut {NoStop}%
\bibitem [{\citenamefont {Magg}\ and\ \citenamefont
  {Wetterich}(1980)}]{Magg:1980ut}%
  \BibitemOpen
  \bibfield  {author} {\bibinfo {author} {\bibfnamefont {M.}~\bibnamefont
  {Magg}}\ and\ \bibinfo {author} {\bibfnamefont {C.}~\bibnamefont
  {Wetterich}},\ }\href {\doibase 10.1016/0370-2693(80)90825-4} {\bibfield
  {journal} {\bibinfo  {journal} {Phys. Lett.}\ }\textbf {\bibinfo {volume}
  {94B}},\ \bibinfo {pages} {61} (\bibinfo {year} {1980})}\BibitemShut
  {NoStop}%
\bibitem [{\citenamefont {Lazarides}\ \emph {et~al.}(1981)\citenamefont
  {Lazarides}, \citenamefont {Shafi},\ and\ \citenamefont
  {Wetterich}}]{Lazarides:1980nt}%
  \BibitemOpen
  \bibfield  {author} {\bibinfo {author} {\bibfnamefont {G.}~\bibnamefont
  {Lazarides}}, \bibinfo {author} {\bibfnamefont {Q.}~\bibnamefont {Shafi}}, \
  and\ \bibinfo {author} {\bibfnamefont {C.}~\bibnamefont {Wetterich}},\ }\href
  {\doibase 10.1016/0550-3213(81)90354-0} {\bibfield  {journal} {\bibinfo
  {journal} {Nucl. Phys.}\ }\textbf {\bibinfo {volume} {B181}},\ \bibinfo
  {pages} {287} (\bibinfo {year} {1981})}\BibitemShut {NoStop}%
\bibitem [{\citenamefont {Chun}\ \emph {et~al.}(2012)\citenamefont {Chun},
  \citenamefont {Lee},\ and\ \citenamefont {Sharma}}]{Chun:2012jw}%
  \BibitemOpen
  \bibfield  {author} {\bibinfo {author} {\bibfnamefont {E.~J.}\ \bibnamefont
  {Chun}}, \bibinfo {author} {\bibfnamefont {H.~M.}\ \bibnamefont {Lee}}, \
  and\ \bibinfo {author} {\bibfnamefont {P.}~\bibnamefont {Sharma}},\ }\href
  {\doibase 10.1007/JHEP11(2012)106} {\bibfield  {journal} {\bibinfo  {journal}
  {JHEP}\ }\textbf {\bibinfo {volume} {11}},\ \bibinfo {pages} {106} (\bibinfo
  {year} {2012})},\ \Eprint {http://arxiv.org/abs/1209.1303} {arXiv:1209.1303
  [hep-ph]} \BibitemShut {NoStop}%
\bibitem [{\citenamefont {Bhupal~Dev}\ \emph {et~al.}(2013)\citenamefont
  {Bhupal~Dev}, \citenamefont {Ghosh}, \citenamefont {Okada},\ and\
  \citenamefont {Saha}}]{Dev:2013ff}%
  \BibitemOpen
  \bibfield  {author} {\bibinfo {author} {\bibfnamefont {P.~S.}\ \bibnamefont
  {Bhupal~Dev}}, \bibinfo {author} {\bibfnamefont {D.~K.}\ \bibnamefont
  {Ghosh}}, \bibinfo {author} {\bibfnamefont {N.}~\bibnamefont {Okada}}, \ and\
  \bibinfo {author} {\bibfnamefont {I.}~\bibnamefont {Saha}},\ }\href {\doibase
  10.1007/JHEP03(2013)150, 10.1007/JHEP05(2013)049} {\bibfield  {journal}
  {\bibinfo  {journal} {JHEP}\ }\textbf {\bibinfo {volume} {03}},\ \bibinfo
  {pages} {150} (\bibinfo {year} {2013})},\ \bibinfo {note} {[Erratum:
  JHEP05,049(2013)]},\ \Eprint {http://arxiv.org/abs/1301.3453}
  {arXiv:1301.3453 [hep-ph]} \BibitemShut {NoStop}%
\bibitem [{\citenamefont {Chakraborty}\ and\ \citenamefont
  {Kundu}(2014)}]{Chakraborty:2014xqa}%
  \BibitemOpen
  \bibfield  {author} {\bibinfo {author} {\bibfnamefont {I.}~\bibnamefont
  {Chakraborty}}\ and\ \bibinfo {author} {\bibfnamefont {A.}~\bibnamefont
  {Kundu}},\ }\href {\doibase 10.1103/PhysRevD.89.095032} {\bibfield  {journal}
  {\bibinfo  {journal} {Phys. Rev.}\ }\textbf {\bibinfo {volume} {D89}},\
  \bibinfo {pages} {095032} (\bibinfo {year} {2014})},\ \Eprint
  {http://arxiv.org/abs/1404.1723} {arXiv:1404.1723 [hep-ph]} \BibitemShut
  {NoStop}%
\bibitem [{\citenamefont {Bennett}\ \emph {et~al.}(2006)\citenamefont {Bennett}
  \emph {et~al.}}]{Muong-2:2006rrc}%
  \BibitemOpen
  \bibfield  {author} {\bibinfo {author} {\bibfnamefont {G.~W.}\ \bibnamefont
  {Bennett}} \emph {et~al.} (\bibinfo {collaboration} {Muon g-2}),\ }\href
  {\doibase 10.1103/PhysRevD.73.072003} {\bibfield  {journal} {\bibinfo
  {journal} {Phys. Rev. D}\ }\textbf {\bibinfo {volume} {73}},\ \bibinfo
  {pages} {072003} (\bibinfo {year} {2006})},\ \Eprint
  {http://arxiv.org/abs/hep-ex/0602035} {arXiv:hep-ex/0602035} \BibitemShut
  {NoStop}%
\bibitem [{\citenamefont {Abi}\ \emph {et~al.}(2021)\citenamefont {Abi} \emph
  {et~al.}}]{Muong-2:2021ojo}%
  \BibitemOpen
  \bibfield  {author} {\bibinfo {author} {\bibfnamefont {B.}~\bibnamefont
  {Abi}} \emph {et~al.} (\bibinfo {collaboration} {Muon g-2}),\ }\href
  {\doibase 10.1103/PhysRevLett.126.141801} {\bibfield  {journal} {\bibinfo
  {journal} {Phys. Rev. Lett.}\ }\textbf {\bibinfo {volume} {126}},\ \bibinfo
  {pages} {141801} (\bibinfo {year} {2021})},\ \Eprint
  {http://arxiv.org/abs/2104.03281} {arXiv:2104.03281 [hep-ex]} \BibitemShut
  {NoStop}%
\bibitem [{\citenamefont {Albahri}\ \emph {et~al.}(2021)\citenamefont {Albahri}
  \emph {et~al.}}]{Muong-2:2021vma}%
  \BibitemOpen
  \bibfield  {author} {\bibinfo {author} {\bibfnamefont {T.}~\bibnamefont
  {Albahri}} \emph {et~al.} (\bibinfo {collaboration} {Muon g-2}),\ }\href
  {\doibase 10.1103/PhysRevD.103.072002} {\bibfield  {journal} {\bibinfo
  {journal} {Phys. Rev. D}\ }\textbf {\bibinfo {volume} {103}},\ \bibinfo
  {pages} {072002} (\bibinfo {year} {2021})},\ \Eprint
  {http://arxiv.org/abs/2104.03247} {arXiv:2104.03247 [hep-ex]} \BibitemShut
  {NoStop}%
\bibitem [{cdf(2022)}]{cdfII:2022}%
  \BibitemOpen
  \href {\doibase 10.1126/science.abk1781} {\bibfield  {journal} {\bibinfo
  {journal} {Science}\ }\textbf {\bibinfo {volume} {376}},\ \bibinfo {pages}
  {170} (\bibinfo {year} {2022})},\ \Eprint
  {http://arxiv.org/abs/https://www.science.org/doi/pdf/10.1126/science.abk1781}
  {https://www.science.org/doi/pdf/10.1126/science.abk1781} \BibitemShut
  {NoStop}%
\bibitem [{\citenamefont {Peskin}\ and\ \citenamefont
  {Takeuchi}(1992)}]{PhysRevD.46.381}%
  \BibitemOpen
  \bibfield  {author} {\bibinfo {author} {\bibfnamefont {M.~E.}\ \bibnamefont
  {Peskin}}\ and\ \bibinfo {author} {\bibfnamefont {T.}~\bibnamefont
  {Takeuchi}},\ }\href {\doibase 10.1103/PhysRevD.46.381} {\bibfield  {journal}
  {\bibinfo  {journal} {Phys. Rev. D}\ }\textbf {\bibinfo {volume} {46}},\
  \bibinfo {pages} {381} (\bibinfo {year} {1992})}\BibitemShut {NoStop}%
\bibitem [{\citenamefont {Kennedy}\ and\ \citenamefont
  {Langacker}(1990)}]{PhysRevLett.65.2967}%
  \BibitemOpen
  \bibfield  {author} {\bibinfo {author} {\bibfnamefont {D.~C.}\ \bibnamefont
  {Kennedy}}\ and\ \bibinfo {author} {\bibfnamefont {P.}~\bibnamefont
  {Langacker}},\ }\href {\doibase 10.1103/PhysRevLett.65.2967} {\bibfield
  {journal} {\bibinfo  {journal} {Phys. Rev. Lett.}\ }\textbf {\bibinfo
  {volume} {65}},\ \bibinfo {pages} {2967} (\bibinfo {year}
  {1990})}\BibitemShut {NoStop}%
\bibitem [{\citenamefont {Strumia}(2022)}]{Strumia:2022qkt}%
  \BibitemOpen
  \bibfield  {author} {\bibinfo {author} {\bibfnamefont {A.}~\bibnamefont
  {Strumia}},\ }\href@noop {} {\  (\bibinfo {year} {2022})},\ \Eprint
  {http://arxiv.org/abs/2204.04191} {arXiv:2204.04191 [hep-ph]} \BibitemShut
  {NoStop}%
\bibitem [{\citenamefont {de~Blas}\ \emph {et~al.}(2022)\citenamefont
  {de~Blas}, \citenamefont {Pierini}, \citenamefont {Reina},\ and\
  \citenamefont {Silvestrini}}]{deBlas:2022hdk}%
  \BibitemOpen
  \bibfield  {author} {\bibinfo {author} {\bibfnamefont {J.}~\bibnamefont
  {de~Blas}}, \bibinfo {author} {\bibfnamefont {M.}~\bibnamefont {Pierini}},
  \bibinfo {author} {\bibfnamefont {L.}~\bibnamefont {Reina}}, \ and\ \bibinfo
  {author} {\bibfnamefont {L.}~\bibnamefont {Silvestrini}},\ }\href@noop {} {\
  (\bibinfo {year} {2022})},\ \Eprint {http://arxiv.org/abs/2204.04204}
  {arXiv:2204.04204 [hep-ph]} \BibitemShut {NoStop}%
\bibitem [{\citenamefont {Paul}\ and\ \citenamefont
  {Valli}(2022)}]{Paul:2022dds}%
  \BibitemOpen
  \bibfield  {author} {\bibinfo {author} {\bibfnamefont {A.}~\bibnamefont
  {Paul}}\ and\ \bibinfo {author} {\bibfnamefont {M.}~\bibnamefont {Valli}},\
  }\href@noop {} {\  (\bibinfo {year} {2022})},\ \Eprint
  {http://arxiv.org/abs/2204.05267} {arXiv:2204.05267 [hep-ph]} \BibitemShut
  {NoStop}%
\bibitem [{\citenamefont {Gu}\ \emph {et~al.}(2022)\citenamefont {Gu},
  \citenamefont {Liu}, \citenamefont {Ma},\ and\ \citenamefont
  {Shu}}]{Gu:2022htv}%
  \BibitemOpen
  \bibfield  {author} {\bibinfo {author} {\bibfnamefont {J.}~\bibnamefont
  {Gu}}, \bibinfo {author} {\bibfnamefont {Z.}~\bibnamefont {Liu}}, \bibinfo
  {author} {\bibfnamefont {T.}~\bibnamefont {Ma}}, \ and\ \bibinfo {author}
  {\bibfnamefont {J.}~\bibnamefont {Shu}},\ }\href@noop {} {\  (\bibinfo {year}
  {2022})},\ \Eprint {http://arxiv.org/abs/2204.05296} {arXiv:2204.05296
  [hep-ph]} \BibitemShut {NoStop}%
\bibitem [{\citenamefont {Asadi}\ \emph {et~al.}(2022)\citenamefont {Asadi},
  \citenamefont {Cesarotti}, \citenamefont {Fraser}, \citenamefont {Homiller},\
  and\ \citenamefont {Parikh}}]{Asadi:2022xiy}%
  \BibitemOpen
  \bibfield  {author} {\bibinfo {author} {\bibfnamefont {P.}~\bibnamefont
  {Asadi}}, \bibinfo {author} {\bibfnamefont {C.}~\bibnamefont {Cesarotti}},
  \bibinfo {author} {\bibfnamefont {K.}~\bibnamefont {Fraser}}, \bibinfo
  {author} {\bibfnamefont {S.}~\bibnamefont {Homiller}}, \ and\ \bibinfo
  {author} {\bibfnamefont {A.}~\bibnamefont {Parikh}},\ }\href@noop {} {\
  (\bibinfo {year} {2022})},\ \Eprint {http://arxiv.org/abs/2204.05283}
  {arXiv:2204.05283 [hep-ph]} \BibitemShut {NoStop}%
\bibitem [{\citenamefont {Endo}\ and\ \citenamefont
  {Mishima}(2022)}]{Endo:2022kiw}%
  \BibitemOpen
  \bibfield  {author} {\bibinfo {author} {\bibfnamefont {M.}~\bibnamefont
  {Endo}}\ and\ \bibinfo {author} {\bibfnamefont {S.}~\bibnamefont {Mishima}},\
  }\href@noop {} {\  (\bibinfo {year} {2022})},\ \Eprint
  {http://arxiv.org/abs/2204.05965} {arXiv:2204.05965 [hep-ph]} \BibitemShut
  {NoStop}%
\bibitem [{\citenamefont {Balkin}\ \emph {et~al.}(2022)\citenamefont {Balkin},
  \citenamefont {Madge}, \citenamefont {Menzo}, \citenamefont {Perez},
  \citenamefont {Soreq},\ and\ \citenamefont {Zupan}}]{Balkin:2022glu}%
  \BibitemOpen
  \bibfield  {author} {\bibinfo {author} {\bibfnamefont {R.}~\bibnamefont
  {Balkin}}, \bibinfo {author} {\bibfnamefont {E.}~\bibnamefont {Madge}},
  \bibinfo {author} {\bibfnamefont {T.}~\bibnamefont {Menzo}}, \bibinfo
  {author} {\bibfnamefont {G.}~\bibnamefont {Perez}}, \bibinfo {author}
  {\bibfnamefont {Y.}~\bibnamefont {Soreq}}, \ and\ \bibinfo {author}
  {\bibfnamefont {J.}~\bibnamefont {Zupan}},\ }\href {\doibase
  10.1007/JHEP05(2022)133} {\bibfield  {journal} {\bibinfo  {journal} {JHEP}\
  }\textbf {\bibinfo {volume} {05}},\ \bibinfo {pages} {133} (\bibinfo {year}
  {2022})},\ \Eprint {http://arxiv.org/abs/2204.05992} {arXiv:2204.05992
  [hep-ph]} \BibitemShut {NoStop}%
\bibitem [{\citenamefont {Yang}\ and\ \citenamefont
  {Zhang}(2022)}]{Yang:2022gvz}%
  \BibitemOpen
  \bibfield  {author} {\bibinfo {author} {\bibfnamefont {J.~M.}\ \bibnamefont
  {Yang}}\ and\ \bibinfo {author} {\bibfnamefont {Y.}~\bibnamefont {Zhang}},\
  }\href@noop {} {\  (\bibinfo {year} {2022})},\ \Eprint
  {http://arxiv.org/abs/2204.04202} {arXiv:2204.04202 [hep-ph]} \BibitemShut
  {NoStop}%
\bibitem [{\citenamefont {Du}\ \emph {et~al.}(2022)\citenamefont {Du},
  \citenamefont {Li}, \citenamefont {Wang},\ and\ \citenamefont
  {Zhang}}]{Du:2022pbp}%
  \BibitemOpen
  \bibfield  {author} {\bibinfo {author} {\bibfnamefont {X.~K.}\ \bibnamefont
  {Du}}, \bibinfo {author} {\bibfnamefont {Z.}~\bibnamefont {Li}}, \bibinfo
  {author} {\bibfnamefont {F.}~\bibnamefont {Wang}}, \ and\ \bibinfo {author}
  {\bibfnamefont {Y.~K.}\ \bibnamefont {Zhang}},\ }\href@noop {} {\  (\bibinfo
  {year} {2022})},\ \Eprint {http://arxiv.org/abs/2204.04286} {arXiv:2204.04286
  [hep-ph]} \BibitemShut {NoStop}%
\bibitem [{\citenamefont {Tang}\ \emph {et~al.}(2022)\citenamefont {Tang},
  \citenamefont {Abdughani}, \citenamefont {Feng}, \citenamefont {Tsai},
  \citenamefont {Wu},\ and\ \citenamefont {Fan}}]{Tang:2022pxh}%
  \BibitemOpen
  \bibfield  {author} {\bibinfo {author} {\bibfnamefont {T.-P.}\ \bibnamefont
  {Tang}}, \bibinfo {author} {\bibfnamefont {M.}~\bibnamefont {Abdughani}},
  \bibinfo {author} {\bibfnamefont {L.}~\bibnamefont {Feng}}, \bibinfo {author}
  {\bibfnamefont {Y.-L.~S.}\ \bibnamefont {Tsai}}, \bibinfo {author}
  {\bibfnamefont {J.}~\bibnamefont {Wu}}, \ and\ \bibinfo {author}
  {\bibfnamefont {Y.-Z.}\ \bibnamefont {Fan}},\ }\href@noop {} {\  (\bibinfo
  {year} {2022})},\ \Eprint {http://arxiv.org/abs/2204.04356} {arXiv:2204.04356
  [hep-ph]} \BibitemShut {NoStop}%
\bibitem [{\citenamefont {Athron}\ \emph
  {et~al.}(2022{\natexlab{a}})\citenamefont {Athron}, \citenamefont {Bach},
  \citenamefont {Jacob}, \citenamefont {Kotlarski}, \citenamefont
  {St\"ockinger},\ and\ \citenamefont {Voigt}}]{Athron:2022isz}%
  \BibitemOpen
  \bibfield  {author} {\bibinfo {author} {\bibfnamefont {P.}~\bibnamefont
  {Athron}}, \bibinfo {author} {\bibfnamefont {M.}~\bibnamefont {Bach}},
  \bibinfo {author} {\bibfnamefont {D.~H.~J.}\ \bibnamefont {Jacob}}, \bibinfo
  {author} {\bibfnamefont {W.}~\bibnamefont {Kotlarski}}, \bibinfo {author}
  {\bibfnamefont {D.}~\bibnamefont {St\"ockinger}}, \ and\ \bibinfo {author}
  {\bibfnamefont {A.}~\bibnamefont {Voigt}},\ }\href@noop {} {\  (\bibinfo
  {year} {2022}{\natexlab{a}})},\ \Eprint {http://arxiv.org/abs/2204.05285}
  {arXiv:2204.05285 [hep-ph]} \BibitemShut {NoStop}%
\bibitem [{\citenamefont {Zheng}\ \emph {et~al.}(2022)\citenamefont {Zheng},
  \citenamefont {Chen},\ and\ \citenamefont {Zhang}}]{Zheng:2022irz}%
  \BibitemOpen
  \bibfield  {author} {\bibinfo {author} {\bibfnamefont {M.-D.}\ \bibnamefont
  {Zheng}}, \bibinfo {author} {\bibfnamefont {F.-Z.}\ \bibnamefont {Chen}}, \
  and\ \bibinfo {author} {\bibfnamefont {H.-H.}\ \bibnamefont {Zhang}},\
  }\href@noop {} {\  (\bibinfo {year} {2022})},\ \Eprint
  {http://arxiv.org/abs/2204.06541} {arXiv:2204.06541 [hep-ph]} \BibitemShut
  {NoStop}%
\bibitem [{\citenamefont {Ghoshal}\ \emph {et~al.}(2022)\citenamefont
  {Ghoshal}, \citenamefont {Okada}, \citenamefont {Okada}, \citenamefont
  {Raut}, \citenamefont {Shafi},\ and\ \citenamefont
  {Thapa}}]{Ghoshal:2022vzo}%
  \BibitemOpen
  \bibfield  {author} {\bibinfo {author} {\bibfnamefont {A.}~\bibnamefont
  {Ghoshal}}, \bibinfo {author} {\bibfnamefont {N.}~\bibnamefont {Okada}},
  \bibinfo {author} {\bibfnamefont {S.}~\bibnamefont {Okada}}, \bibinfo
  {author} {\bibfnamefont {D.}~\bibnamefont {Raut}}, \bibinfo {author}
  {\bibfnamefont {Q.}~\bibnamefont {Shafi}}, \ and\ \bibinfo {author}
  {\bibfnamefont {A.}~\bibnamefont {Thapa}},\ }\href@noop {} {\  (\bibinfo
  {year} {2022})},\ \Eprint {http://arxiv.org/abs/2204.07138} {arXiv:2204.07138
  [hep-ph]} \BibitemShut {NoStop}%
\bibitem [{\citenamefont {Lu}\ \emph {et~al.}(2022)\citenamefont {Lu},
  \citenamefont {Wu}, \citenamefont {Wu},\ and\ \citenamefont
  {Zhu}}]{Lu:2022bgw}%
  \BibitemOpen
  \bibfield  {author} {\bibinfo {author} {\bibfnamefont {C.-T.}\ \bibnamefont
  {Lu}}, \bibinfo {author} {\bibfnamefont {L.}~\bibnamefont {Wu}}, \bibinfo
  {author} {\bibfnamefont {Y.}~\bibnamefont {Wu}}, \ and\ \bibinfo {author}
  {\bibfnamefont {B.}~\bibnamefont {Zhu}},\ }\href@noop {} {\  (\bibinfo {year}
  {2022})},\ \Eprint {http://arxiv.org/abs/2204.03796} {arXiv:2204.03796
  [hep-ph]} \BibitemShut {NoStop}%
\bibitem [{\citenamefont {Fan}\ \emph {et~al.}(2022)\citenamefont {Fan},
  \citenamefont {Tang}, \citenamefont {Tsai},\ and\ \citenamefont
  {Wu}}]{Fan:2022dck}%
  \BibitemOpen
  \bibfield  {author} {\bibinfo {author} {\bibfnamefont {Y.-Z.}\ \bibnamefont
  {Fan}}, \bibinfo {author} {\bibfnamefont {T.-P.}\ \bibnamefont {Tang}},
  \bibinfo {author} {\bibfnamefont {Y.-L.~S.}\ \bibnamefont {Tsai}}, \ and\
  \bibinfo {author} {\bibfnamefont {L.}~\bibnamefont {Wu}},\ }\href@noop {} {\
  (\bibinfo {year} {2022})},\ \Eprint {http://arxiv.org/abs/2204.03693}
  {arXiv:2204.03693 [hep-ph]} \BibitemShut {NoStop}%
\bibitem [{\citenamefont {Zhu}\ \emph {et~al.}(2022)\citenamefont {Zhu},
  \citenamefont {Li}, \citenamefont {Cheng}, \citenamefont {Li},\ and\
  \citenamefont {Liang}}]{Zhu:2022scj}%
  \BibitemOpen
  \bibfield  {author} {\bibinfo {author} {\bibfnamefont {B.-Y.}\ \bibnamefont
  {Zhu}}, \bibinfo {author} {\bibfnamefont {S.}~\bibnamefont {Li}}, \bibinfo
  {author} {\bibfnamefont {J.-G.}\ \bibnamefont {Cheng}}, \bibinfo {author}
  {\bibfnamefont {R.-L.}\ \bibnamefont {Li}}, \ and\ \bibinfo {author}
  {\bibfnamefont {Y.-F.}\ \bibnamefont {Liang}},\ }\href@noop {} {\  (\bibinfo
  {year} {2022})},\ \Eprint {http://arxiv.org/abs/2204.04688} {arXiv:2204.04688
  [astro-ph.HE]} \BibitemShut {NoStop}%
\bibitem [{\citenamefont {Song}\ \emph {et~al.}(2022)\citenamefont {Song},
  \citenamefont {Su},\ and\ \citenamefont {Zhang}}]{Song:2022xts}%
  \BibitemOpen
  \bibfield  {author} {\bibinfo {author} {\bibfnamefont {H.}~\bibnamefont
  {Song}}, \bibinfo {author} {\bibfnamefont {W.}~\bibnamefont {Su}}, \ and\
  \bibinfo {author} {\bibfnamefont {M.}~\bibnamefont {Zhang}},\ }\href@noop {}
  {\  (\bibinfo {year} {2022})},\ \Eprint {http://arxiv.org/abs/2204.05085}
  {arXiv:2204.05085 [hep-ph]} \BibitemShut {NoStop}%
\bibitem [{\citenamefont {Mondal}(2022)}]{Mondal:2022xdy}%
  \BibitemOpen
  \bibfield  {author} {\bibinfo {author} {\bibfnamefont {P.}~\bibnamefont
  {Mondal}},\ }\href@noop {} {\  (\bibinfo {year} {2022})},\ \Eprint
  {http://arxiv.org/abs/2204.07844} {arXiv:2204.07844 [hep-ph]} \BibitemShut
  {NoStop}%
\bibitem [{\citenamefont {Ghosh}\ \emph {et~al.}(2022)\citenamefont {Ghosh},
  \citenamefont {Mukhopadhyaya},\ and\ \citenamefont {Sarkar}}]{Ghosh:2022zqs}%
  \BibitemOpen
  \bibfield  {author} {\bibinfo {author} {\bibfnamefont {R.}~\bibnamefont
  {Ghosh}}, \bibinfo {author} {\bibfnamefont {B.}~\bibnamefont
  {Mukhopadhyaya}}, \ and\ \bibinfo {author} {\bibfnamefont {U.}~\bibnamefont
  {Sarkar}},\ }\href@noop {} {\  (\bibinfo {year} {2022})},\ \Eprint
  {http://arxiv.org/abs/2205.05041} {arXiv:2205.05041 [hep-ph]} \BibitemShut
  {NoStop}%
\bibitem [{\citenamefont {Bahl}\ \emph {et~al.}(2022)\citenamefont {Bahl},
  \citenamefont {Braathen},\ and\ \citenamefont {Weiglein}}]{Bahl:2022xzi}%
  \BibitemOpen
  \bibfield  {author} {\bibinfo {author} {\bibfnamefont {H.}~\bibnamefont
  {Bahl}}, \bibinfo {author} {\bibfnamefont {J.}~\bibnamefont {Braathen}}, \
  and\ \bibinfo {author} {\bibfnamefont {G.}~\bibnamefont {Weiglein}},\
  }\href@noop {} {\  (\bibinfo {year} {2022})},\ \Eprint
  {http://arxiv.org/abs/2204.05269} {arXiv:2204.05269 [hep-ph]} \BibitemShut
  {NoStop}%
\bibitem [{\citenamefont {Heo}\ \emph {et~al.}(2022)\citenamefont {Heo},
  \citenamefont {Jung},\ and\ \citenamefont {Lee}}]{Heo:2022dey}%
  \BibitemOpen
  \bibfield  {author} {\bibinfo {author} {\bibfnamefont {Y.}~\bibnamefont
  {Heo}}, \bibinfo {author} {\bibfnamefont {D.-W.}\ \bibnamefont {Jung}}, \
  and\ \bibinfo {author} {\bibfnamefont {J.~S.}\ \bibnamefont {Lee}},\
  }\href@noop {} {\  (\bibinfo {year} {2022})},\ \Eprint
  {http://arxiv.org/abs/2204.05728} {arXiv:2204.05728 [hep-ph]} \BibitemShut
  {NoStop}%
\bibitem [{\citenamefont {Babu}\ \emph {et~al.}(2022)\citenamefont {Babu},
  \citenamefont {Jana},\ and\ \citenamefont {K.}}]{Babu:2022pdn}%
  \BibitemOpen
  \bibfield  {author} {\bibinfo {author} {\bibfnamefont {K.~S.}\ \bibnamefont
  {Babu}}, \bibinfo {author} {\bibfnamefont {S.}~\bibnamefont {Jana}}, \ and\
  \bibinfo {author} {\bibfnamefont {V.~P.}\ \bibnamefont {K.}},\ }\href@noop {}
  {\  (\bibinfo {year} {2022})},\ \Eprint {http://arxiv.org/abs/2204.05303}
  {arXiv:2204.05303 [hep-ph]} \BibitemShut {NoStop}%
\bibitem [{\citenamefont {Biek\"otter}\ \emph {et~al.}(2022)\citenamefont
  {Biek\"otter}, \citenamefont {Heinemeyer},\ and\ \citenamefont
  {Weiglein}}]{Biekotter:2022abc}%
  \BibitemOpen
  \bibfield  {author} {\bibinfo {author} {\bibfnamefont {T.}~\bibnamefont
  {Biek\"otter}}, \bibinfo {author} {\bibfnamefont {S.}~\bibnamefont
  {Heinemeyer}}, \ and\ \bibinfo {author} {\bibfnamefont {G.}~\bibnamefont
  {Weiglein}},\ }\href@noop {} {\  (\bibinfo {year} {2022})},\ \Eprint
  {http://arxiv.org/abs/2204.05975} {arXiv:2204.05975 [hep-ph]} \BibitemShut
  {NoStop}%
\bibitem [{\citenamefont {Ahn}\ \emph {et~al.}(2022)\citenamefont {Ahn},
  \citenamefont {Kang},\ and\ \citenamefont {Ramos}}]{Ahn:2022xeq}%
  \BibitemOpen
  \bibfield  {author} {\bibinfo {author} {\bibfnamefont {Y.~H.}\ \bibnamefont
  {Ahn}}, \bibinfo {author} {\bibfnamefont {S.~K.}\ \bibnamefont {Kang}}, \
  and\ \bibinfo {author} {\bibfnamefont {R.}~\bibnamefont {Ramos}},\
  }\href@noop {} {\  (\bibinfo {year} {2022})},\ \Eprint
  {http://arxiv.org/abs/2204.06485} {arXiv:2204.06485 [hep-ph]} \BibitemShut
  {NoStop}%
\bibitem [{\citenamefont {Han}\ \emph {et~al.}(2022)\citenamefont {Han},
  \citenamefont {Wang}, \citenamefont {Wang}, \citenamefont {Yang},\ and\
  \citenamefont {Zhang}}]{Han:2022juu}%
  \BibitemOpen
  \bibfield  {author} {\bibinfo {author} {\bibfnamefont {X.-F.}\ \bibnamefont
  {Han}}, \bibinfo {author} {\bibfnamefont {F.}~\bibnamefont {Wang}}, \bibinfo
  {author} {\bibfnamefont {L.}~\bibnamefont {Wang}}, \bibinfo {author}
  {\bibfnamefont {J.~M.}\ \bibnamefont {Yang}}, \ and\ \bibinfo {author}
  {\bibfnamefont {Y.}~\bibnamefont {Zhang}},\ }\href@noop {} {\  (\bibinfo
  {year} {2022})},\ \Eprint {http://arxiv.org/abs/2204.06505} {arXiv:2204.06505
  [hep-ph]} \BibitemShut {NoStop}%
\bibitem [{\citenamefont {Arcadi}\ and\ \citenamefont
  {Djouadi}(2022)}]{Arcadi:2022dmt}%
  \BibitemOpen
  \bibfield  {author} {\bibinfo {author} {\bibfnamefont {G.}~\bibnamefont
  {Arcadi}}\ and\ \bibinfo {author} {\bibfnamefont {A.}~\bibnamefont
  {Djouadi}},\ }\href@noop {} {\  (\bibinfo {year} {2022})},\ \Eprint
  {http://arxiv.org/abs/2204.08406} {arXiv:2204.08406 [hep-ph]} \BibitemShut
  {NoStop}%
\bibitem [{\citenamefont {Lee}\ \emph {et~al.}(2022)\citenamefont {Lee},
  \citenamefont {Cheung}, \citenamefont {Kim}, \citenamefont {Lu},\ and\
  \citenamefont {Song}}]{Lee:2022gyf}%
  \BibitemOpen
  \bibfield  {author} {\bibinfo {author} {\bibfnamefont {S.}~\bibnamefont
  {Lee}}, \bibinfo {author} {\bibfnamefont {K.}~\bibnamefont {Cheung}},
  \bibinfo {author} {\bibfnamefont {J.}~\bibnamefont {Kim}}, \bibinfo {author}
  {\bibfnamefont {C.-T.}\ \bibnamefont {Lu}}, \ and\ \bibinfo {author}
  {\bibfnamefont {J.}~\bibnamefont {Song}},\ }\href@noop {} {\  (\bibinfo
  {year} {2022})},\ \Eprint {http://arxiv.org/abs/2204.10338} {arXiv:2204.10338
  [hep-ph]} \BibitemShut {NoStop}%
\bibitem [{\citenamefont {Ghorbani}\ and\ \citenamefont
  {Ghorbani}(2022)}]{Ghorbani:2022vtv}%
  \BibitemOpen
  \bibfield  {author} {\bibinfo {author} {\bibfnamefont {K.}~\bibnamefont
  {Ghorbani}}\ and\ \bibinfo {author} {\bibfnamefont {P.}~\bibnamefont
  {Ghorbani}},\ }\href@noop {} {\  (\bibinfo {year} {2022})},\ \Eprint
  {http://arxiv.org/abs/2204.09001} {arXiv:2204.09001 [hep-ph]} \BibitemShut
  {NoStop}%
\bibitem [{\citenamefont {Batra}\ \emph
  {et~al.}(2022{\natexlab{a}})\citenamefont {Batra}, \citenamefont {A},
  \citenamefont {Mandal}, \citenamefont {Prajapati},\ and\ \citenamefont
  {Srivastava}}]{Batra:2022pej}%
  \BibitemOpen
  \bibfield  {author} {\bibinfo {author} {\bibfnamefont {A.}~\bibnamefont
  {Batra}}, \bibinfo {author} {\bibfnamefont {S.~K.}\ \bibnamefont {A}},
  \bibinfo {author} {\bibfnamefont {S.}~\bibnamefont {Mandal}}, \bibinfo
  {author} {\bibfnamefont {H.}~\bibnamefont {Prajapati}}, \ and\ \bibinfo
  {author} {\bibfnamefont {R.}~\bibnamefont {Srivastava}},\ }\href@noop {} {\
  (\bibinfo {year} {2022}{\natexlab{a}})},\ \Eprint
  {http://arxiv.org/abs/2204.11945} {arXiv:2204.11945 [hep-ph]} \BibitemShut
  {NoStop}%
\bibitem [{\citenamefont {Batra}\ \emph
  {et~al.}(2022{\natexlab{b}})\citenamefont {Batra}, \citenamefont {K.~A.},
  \citenamefont {Mandal},\ and\ \citenamefont {Srivastava}}]{Batra:2022org}%
  \BibitemOpen
  \bibfield  {author} {\bibinfo {author} {\bibfnamefont {A.}~\bibnamefont
  {Batra}}, \bibinfo {author} {\bibfnamefont {S.}~\bibnamefont {K.~A.}},
  \bibinfo {author} {\bibfnamefont {S.}~\bibnamefont {Mandal}}, \ and\ \bibinfo
  {author} {\bibfnamefont {R.}~\bibnamefont {Srivastava}},\ }\href@noop {} {\
  (\bibinfo {year} {2022}{\natexlab{b}})},\ \Eprint
  {http://arxiv.org/abs/2204.09376} {arXiv:2204.09376 [hep-ph]} \BibitemShut
  {NoStop}%
\bibitem [{\citenamefont {Popov}\ and\ \citenamefont
  {Srivastava}(2022)}]{Popov:2022ldh}%
  \BibitemOpen
  \bibfield  {author} {\bibinfo {author} {\bibfnamefont {O.}~\bibnamefont
  {Popov}}\ and\ \bibinfo {author} {\bibfnamefont {R.}~\bibnamefont
  {Srivastava}},\ }\href@noop {} {\  (\bibinfo {year} {2022})},\ \Eprint
  {http://arxiv.org/abs/2204.08568} {arXiv:2204.08568 [hep-ph]} \BibitemShut
  {NoStop}%
\bibitem [{\citenamefont {Lee}\ and\ \citenamefont
  {Yamashita}(2022)}]{Lee:2022nqz}%
  \BibitemOpen
  \bibfield  {author} {\bibinfo {author} {\bibfnamefont {H.~M.}\ \bibnamefont
  {Lee}}\ and\ \bibinfo {author} {\bibfnamefont {K.}~\bibnamefont
  {Yamashita}},\ }\href@noop {} {\  (\bibinfo {year} {2022})},\ \Eprint
  {http://arxiv.org/abs/2204.05024} {arXiv:2204.05024 [hep-ph]} \BibitemShut
  {NoStop}%
\bibitem [{\citenamefont {Kim}\ \emph {et~al.}(2022)\citenamefont {Kim},
  \citenamefont {Lee}, \citenamefont {Menkara},\ and\ \citenamefont
  {Yamashita}}]{Kim:2022zhj}%
  \BibitemOpen
  \bibfield  {author} {\bibinfo {author} {\bibfnamefont {S.-S.}\ \bibnamefont
  {Kim}}, \bibinfo {author} {\bibfnamefont {H.~M.}\ \bibnamefont {Lee}},
  \bibinfo {author} {\bibfnamefont {A.~G.}\ \bibnamefont {Menkara}}, \ and\
  \bibinfo {author} {\bibfnamefont {K.}~\bibnamefont {Yamashita}},\ }\href@noop
  {} {\  (\bibinfo {year} {2022})},\ \Eprint {http://arxiv.org/abs/2205.04016}
  {arXiv:2205.04016 [hep-ph]} \BibitemShut {NoStop}%
\bibitem [{\citenamefont {Kawamura}\ \emph {et~al.}(2022)\citenamefont
  {Kawamura}, \citenamefont {Okawa},\ and\ \citenamefont
  {Omura}}]{Kawamura:2022uft}%
  \BibitemOpen
  \bibfield  {author} {\bibinfo {author} {\bibfnamefont {J.}~\bibnamefont
  {Kawamura}}, \bibinfo {author} {\bibfnamefont {S.}~\bibnamefont {Okawa}}, \
  and\ \bibinfo {author} {\bibfnamefont {Y.}~\bibnamefont {Omura}},\
  }\href@noop {} {\  (\bibinfo {year} {2022})},\ \Eprint
  {http://arxiv.org/abs/2204.07022} {arXiv:2204.07022 [hep-ph]} \BibitemShut
  {NoStop}%
\bibitem [{\citenamefont {Crivellin}\ \emph {et~al.}(2022)\citenamefont
  {Crivellin}, \citenamefont {Kirk}, \citenamefont {Kitahara},\ and\
  \citenamefont {Mescia}}]{Crivellin:2022fdf}%
  \BibitemOpen
  \bibfield  {author} {\bibinfo {author} {\bibfnamefont {A.}~\bibnamefont
  {Crivellin}}, \bibinfo {author} {\bibfnamefont {M.}~\bibnamefont {Kirk}},
  \bibinfo {author} {\bibfnamefont {T.}~\bibnamefont {Kitahara}}, \ and\
  \bibinfo {author} {\bibfnamefont {F.}~\bibnamefont {Mescia}},\ }\href@noop {}
  {\  (\bibinfo {year} {2022})},\ \Eprint {http://arxiv.org/abs/2204.05962}
  {arXiv:2204.05962 [hep-ph]} \BibitemShut {NoStop}%
\bibitem [{\citenamefont {Nagao}\ \emph {et~al.}(2022)\citenamefont {Nagao},
  \citenamefont {Nomura},\ and\ \citenamefont {Okada}}]{Nagao:2022oin}%
  \BibitemOpen
  \bibfield  {author} {\bibinfo {author} {\bibfnamefont {K.~I.}\ \bibnamefont
  {Nagao}}, \bibinfo {author} {\bibfnamefont {T.}~\bibnamefont {Nomura}}, \
  and\ \bibinfo {author} {\bibfnamefont {H.}~\bibnamefont {Okada}},\
  }\href@noop {} {\  (\bibinfo {year} {2022})},\ \Eprint
  {http://arxiv.org/abs/2204.07411} {arXiv:2204.07411 [hep-ph]} \BibitemShut
  {NoStop}%
\bibitem [{\citenamefont {Bhaskar}\ \emph {et~al.}(2022)\citenamefont
  {Bhaskar}, \citenamefont {Madathil}, \citenamefont {Mandal},\ and\
  \citenamefont {Mitra}}]{Bhaskar:2022vgk}%
  \BibitemOpen
  \bibfield  {author} {\bibinfo {author} {\bibfnamefont {A.}~\bibnamefont
  {Bhaskar}}, \bibinfo {author} {\bibfnamefont {A.~A.}\ \bibnamefont
  {Madathil}}, \bibinfo {author} {\bibfnamefont {T.}~\bibnamefont {Mandal}}, \
  and\ \bibinfo {author} {\bibfnamefont {S.}~\bibnamefont {Mitra}},\
  }\href@noop {} {\  (\bibinfo {year} {2022})},\ \Eprint
  {http://arxiv.org/abs/2204.09031} {arXiv:2204.09031 [hep-ph]} \BibitemShut
  {NoStop}%
\bibitem [{\citenamefont {Cheung}\ \emph {et~al.}(2022)\citenamefont {Cheung},
  \citenamefont {Keung},\ and\ \citenamefont {Tseng}}]{Cheung:2022zsb}%
  \BibitemOpen
  \bibfield  {author} {\bibinfo {author} {\bibfnamefont {K.}~\bibnamefont
  {Cheung}}, \bibinfo {author} {\bibfnamefont {W.-Y.}\ \bibnamefont {Keung}}, \
  and\ \bibinfo {author} {\bibfnamefont {P.-Y.}\ \bibnamefont {Tseng}},\
  }\href@noop {} {\  (\bibinfo {year} {2022})},\ \Eprint
  {http://arxiv.org/abs/2204.05942} {arXiv:2204.05942 [hep-ph]} \BibitemShut
  {NoStop}%
\bibitem [{\citenamefont {Athron}\ \emph
  {et~al.}(2022{\natexlab{b}})\citenamefont {Athron}, \citenamefont {Fowlie},
  \citenamefont {Lu}, \citenamefont {Wu}, \citenamefont {Wu},\ and\
  \citenamefont {Zhu}}]{Athron:2022qpo}%
  \BibitemOpen
  \bibfield  {author} {\bibinfo {author} {\bibfnamefont {P.}~\bibnamefont
  {Athron}}, \bibinfo {author} {\bibfnamefont {A.}~\bibnamefont {Fowlie}},
  \bibinfo {author} {\bibfnamefont {C.-T.}\ \bibnamefont {Lu}}, \bibinfo
  {author} {\bibfnamefont {L.}~\bibnamefont {Wu}}, \bibinfo {author}
  {\bibfnamefont {Y.}~\bibnamefont {Wu}}, \ and\ \bibinfo {author}
  {\bibfnamefont {B.}~\bibnamefont {Zhu}},\ }\href@noop {} {\  (\bibinfo {year}
  {2022}{\natexlab{b}})},\ \Eprint {http://arxiv.org/abs/2204.03996}
  {arXiv:2204.03996 [hep-ph]} \BibitemShut {NoStop}%
\bibitem [{\citenamefont {Bagnaschi}\ \emph {et~al.}(2022)\citenamefont
  {Bagnaschi}, \citenamefont {Ellis}, \citenamefont {Madigan}, \citenamefont
  {Mimasu}, \citenamefont {Sanz},\ and\ \citenamefont
  {You}}]{Bagnaschi:2022whn}%
  \BibitemOpen
  \bibfield  {author} {\bibinfo {author} {\bibfnamefont {E.}~\bibnamefont
  {Bagnaschi}}, \bibinfo {author} {\bibfnamefont {J.}~\bibnamefont {Ellis}},
  \bibinfo {author} {\bibfnamefont {M.}~\bibnamefont {Madigan}}, \bibinfo
  {author} {\bibfnamefont {K.}~\bibnamefont {Mimasu}}, \bibinfo {author}
  {\bibfnamefont {V.}~\bibnamefont {Sanz}}, \ and\ \bibinfo {author}
  {\bibfnamefont {T.}~\bibnamefont {You}},\ }\href@noop {} {\  (\bibinfo {year}
  {2022})},\ \Eprint {http://arxiv.org/abs/2204.05260} {arXiv:2204.05260
  [hep-ph]} \BibitemShut {NoStop}%
\bibitem [{\citenamefont {Di~Luzio}\ \emph {et~al.}(2022)\citenamefont
  {Di~Luzio}, \citenamefont {Gr\"ober},\ and\ \citenamefont
  {Paradisi}}]{DiLuzio:2022xns}%
  \BibitemOpen
  \bibfield  {author} {\bibinfo {author} {\bibfnamefont {L.}~\bibnamefont
  {Di~Luzio}}, \bibinfo {author} {\bibfnamefont {R.}~\bibnamefont {Gr\"ober}},
  \ and\ \bibinfo {author} {\bibfnamefont {P.}~\bibnamefont {Paradisi}},\
  }\href@noop {} {\  (\bibinfo {year} {2022})},\ \Eprint
  {http://arxiv.org/abs/2204.05284} {arXiv:2204.05284 [hep-ph]} \BibitemShut
  {NoStop}%
\bibitem [{\citenamefont {Cirigliano}\ \emph {et~al.}(2022)\citenamefont
  {Cirigliano}, \citenamefont {Dekens}, \citenamefont {de~Vries}, \citenamefont
  {Mereghetti},\ and\ \citenamefont {Tong}}]{Cirigliano:2022qdm}%
  \BibitemOpen
  \bibfield  {author} {\bibinfo {author} {\bibfnamefont {V.}~\bibnamefont
  {Cirigliano}}, \bibinfo {author} {\bibfnamefont {W.}~\bibnamefont {Dekens}},
  \bibinfo {author} {\bibfnamefont {J.}~\bibnamefont {de~Vries}}, \bibinfo
  {author} {\bibfnamefont {E.}~\bibnamefont {Mereghetti}}, \ and\ \bibinfo
  {author} {\bibfnamefont {T.}~\bibnamefont {Tong}},\ }\href@noop {} {\
  (\bibinfo {year} {2022})},\ \Eprint {http://arxiv.org/abs/2204.08440}
  {arXiv:2204.08440 [hep-ph]} \BibitemShut {NoStop}%
\bibitem [{\citenamefont {Gupta}(2022)}]{Gupta:2022lrt}%
  \BibitemOpen
  \bibfield  {author} {\bibinfo {author} {\bibfnamefont {R.~S.}\ \bibnamefont
  {Gupta}},\ }\href@noop {} {\  (\bibinfo {year} {2022})},\ \Eprint
  {http://arxiv.org/abs/2204.13690} {arXiv:2204.13690 [hep-ph]} \BibitemShut
  {NoStop}%
\bibitem [{\citenamefont {Fukuyama}\ \emph {et~al.}(2010)\citenamefont
  {Fukuyama}, \citenamefont {Sugiyama},\ and\ \citenamefont
  {Tsumura}}]{Fukuyama:2009xk}%
  \BibitemOpen
  \bibfield  {author} {\bibinfo {author} {\bibfnamefont {T.}~\bibnamefont
  {Fukuyama}}, \bibinfo {author} {\bibfnamefont {H.}~\bibnamefont {Sugiyama}},
  \ and\ \bibinfo {author} {\bibfnamefont {K.}~\bibnamefont {Tsumura}},\ }\href
  {\doibase 10.1007/JHEP03(2010)044} {\bibfield  {journal} {\bibinfo  {journal}
  {JHEP}\ }\textbf {\bibinfo {volume} {03}},\ \bibinfo {pages} {044} (\bibinfo
  {year} {2010})},\ \Eprint {http://arxiv.org/abs/0909.4943} {arXiv:0909.4943
  [hep-ph]} \BibitemShut {NoStop}%
\bibitem [{\citenamefont {Chakrabarty}(2021)}]{Chakrabarty:2020jro}%
  \BibitemOpen
  \bibfield  {author} {\bibinfo {author} {\bibfnamefont {N.}~\bibnamefont
  {Chakrabarty}},\ }\href {\doibase 10.1140/epjp/s13360-021-02168-3} {\bibfield
   {journal} {\bibinfo  {journal} {Eur. Phys. J. Plus}\ }\textbf {\bibinfo
  {volume} {136}},\ \bibinfo {pages} {1183} (\bibinfo {year} {2021})},\ \Eprint
  {http://arxiv.org/abs/2010.05215} {arXiv:2010.05215 [hep-ph]} \BibitemShut
  {NoStop}%
\bibitem [{\citenamefont {Thomas}\ and\ \citenamefont
  {Wells}(1998)}]{Thomas:1998wy}%
  \BibitemOpen
  \bibfield  {author} {\bibinfo {author} {\bibfnamefont {S.~D.}\ \bibnamefont
  {Thomas}}\ and\ \bibinfo {author} {\bibfnamefont {J.~D.}\ \bibnamefont
  {Wells}},\ }\href {\doibase 10.1103/PhysRevLett.81.34} {\bibfield  {journal}
  {\bibinfo  {journal} {Phys. Rev. Lett.}\ }\textbf {\bibinfo {volume} {81}},\
  \bibinfo {pages} {34} (\bibinfo {year} {1998})},\ \Eprint
  {http://arxiv.org/abs/hep-ph/9804359} {arXiv:hep-ph/9804359} \BibitemShut
  {NoStop}%
\bibitem [{\citenamefont {Freitas}\ \emph {et~al.}(2021)\citenamefont
  {Freitas}, \citenamefont {Gon\c{c}alves}, \citenamefont {Morais},\ and\
  \citenamefont {Pasechnik}}]{Freitas:2020ttd}%
  \BibitemOpen
  \bibfield  {author} {\bibinfo {author} {\bibfnamefont {F.~F.}\ \bibnamefont
  {Freitas}}, \bibinfo {author} {\bibfnamefont {J.~a.}\ \bibnamefont
  {Gon\c{c}alves}}, \bibinfo {author} {\bibfnamefont {A.~P.}\ \bibnamefont
  {Morais}}, \ and\ \bibinfo {author} {\bibfnamefont {R.}~\bibnamefont
  {Pasechnik}},\ }\href {\doibase 10.1007/JHEP01(2021)076} {\bibfield
  {journal} {\bibinfo  {journal} {JHEP}\ }\textbf {\bibinfo {volume} {01}},\
  \bibinfo {pages} {076} (\bibinfo {year} {2021})},\ \Eprint
  {http://arxiv.org/abs/2010.01307} {arXiv:2010.01307 [hep-ph]} \BibitemShut
  {NoStop}%
\bibitem [{\citenamefont {Kannike}\ \emph {et~al.}(2012)\citenamefont
  {Kannike}, \citenamefont {Raidal}, \citenamefont {Straub},\ and\
  \citenamefont {Strumia}}]{Kannike:2011ng}%
  \BibitemOpen
  \bibfield  {author} {\bibinfo {author} {\bibfnamefont {K.}~\bibnamefont
  {Kannike}}, \bibinfo {author} {\bibfnamefont {M.}~\bibnamefont {Raidal}},
  \bibinfo {author} {\bibfnamefont {D.~M.}\ \bibnamefont {Straub}}, \ and\
  \bibinfo {author} {\bibfnamefont {A.}~\bibnamefont {Strumia}},\ }\href
  {\doibase 10.1007/JHEP02(2012)106} {\bibfield  {journal} {\bibinfo  {journal}
  {JHEP}\ }\textbf {\bibinfo {volume} {02}},\ \bibinfo {pages} {106} (\bibinfo
  {year} {2012})},\ \bibinfo {note} {[Erratum: JHEP 10, 136 (2012)]},\ \Eprint
  {http://arxiv.org/abs/1111.2551} {arXiv:1111.2551 [hep-ph]} \BibitemShut
  {NoStop}%
\bibitem [{\citenamefont {Dermisek}\ and\ \citenamefont
  {Raval}(2013)}]{Dermisek:2013gta}%
  \BibitemOpen
  \bibfield  {author} {\bibinfo {author} {\bibfnamefont {R.}~\bibnamefont
  {Dermisek}}\ and\ \bibinfo {author} {\bibfnamefont {A.}~\bibnamefont
  {Raval}},\ }\href {\doibase 10.1103/PhysRevD.88.013017} {\bibfield  {journal}
  {\bibinfo  {journal} {Phys. Rev. D}\ }\textbf {\bibinfo {volume} {88}},\
  \bibinfo {pages} {013017} (\bibinfo {year} {2013})},\ \Eprint
  {http://arxiv.org/abs/1305.3522} {arXiv:1305.3522 [hep-ph]} \BibitemShut
  {NoStop}%
\bibitem [{\citenamefont {Megias}\ \emph {et~al.}(2017)\citenamefont {Megias},
  \citenamefont {Quiros},\ and\ \citenamefont {Salas}}]{Megias:2017dzd}%
  \BibitemOpen
  \bibfield  {author} {\bibinfo {author} {\bibfnamefont {E.}~\bibnamefont
  {Megias}}, \bibinfo {author} {\bibfnamefont {M.}~\bibnamefont {Quiros}}, \
  and\ \bibinfo {author} {\bibfnamefont {L.}~\bibnamefont {Salas}},\ }\href
  {\doibase 10.1007/JHEP05(2017)016} {\bibfield  {journal} {\bibinfo  {journal}
  {JHEP}\ }\textbf {\bibinfo {volume} {05}},\ \bibinfo {pages} {016} (\bibinfo
  {year} {2017})},\ \Eprint {http://arxiv.org/abs/1701.05072} {arXiv:1701.05072
  [hep-ph]} \BibitemShut {NoStop}%
\bibitem [{\citenamefont {Crivellin}\ \emph {et~al.}(2018)\citenamefont
  {Crivellin}, \citenamefont {Hoferichter},\ and\ \citenamefont
  {Schmidt-Wellenburg}}]{Crivellin:2018qmi}%
  \BibitemOpen
  \bibfield  {author} {\bibinfo {author} {\bibfnamefont {A.}~\bibnamefont
  {Crivellin}}, \bibinfo {author} {\bibfnamefont {M.}~\bibnamefont
  {Hoferichter}}, \ and\ \bibinfo {author} {\bibfnamefont {P.}~\bibnamefont
  {Schmidt-Wellenburg}},\ }\href {\doibase 10.1103/PhysRevD.98.113002}
  {\bibfield  {journal} {\bibinfo  {journal} {Phys. Rev. D}\ }\textbf {\bibinfo
  {volume} {98}},\ \bibinfo {pages} {113002} (\bibinfo {year} {2018})},\
  \Eprint {http://arxiv.org/abs/1807.11484} {arXiv:1807.11484 [hep-ph]}
  \BibitemShut {NoStop}%
\bibitem [{\citenamefont {Aad}\ \emph {et~al.}(2020)\citenamefont {Aad} \emph
  {et~al.}}]{Aad:2020xfq}%
  \BibitemOpen
  \bibfield  {author} {\bibinfo {author} {\bibfnamefont {G.}~\bibnamefont
  {Aad}} \emph {et~al.} (\bibinfo {collaboration} {ATLAS}),\ }\href@noop {} {\
  (\bibinfo {year} {2020})},\ \Eprint {http://arxiv.org/abs/2007.07830}
  {arXiv:2007.07830 [hep-ex]} \BibitemShut {NoStop}%
\bibitem [{\citenamefont {Sirunyan}\ \emph {et~al.}(2019)\citenamefont
  {Sirunyan} \emph {et~al.}}]{Sirunyan:2018hbu}%
  \BibitemOpen
  \bibfield  {author} {\bibinfo {author} {\bibfnamefont {A.~M.}\ \bibnamefont
  {Sirunyan}} \emph {et~al.} (\bibinfo {collaboration} {CMS}),\ }\href
  {\doibase 10.1103/PhysRevLett.122.021801} {\bibfield  {journal} {\bibinfo
  {journal} {Phys. Rev. Lett.}\ }\textbf {\bibinfo {volume} {122}},\ \bibinfo
  {pages} {021801} (\bibinfo {year} {2019})},\ \Eprint
  {http://arxiv.org/abs/1807.06325} {arXiv:1807.06325 [hep-ex]} \BibitemShut
  {NoStop}%
\bibitem [{\citenamefont {Frank}\ and\ \citenamefont
  {Saha}(2020)}]{Frank:2020smf}%
  \BibitemOpen
  \bibfield  {author} {\bibinfo {author} {\bibfnamefont {M.}~\bibnamefont
  {Frank}}\ and\ \bibinfo {author} {\bibfnamefont {I.}~\bibnamefont {Saha}},\
  }\href@noop {} {\  (\bibinfo {year} {2020})},\ \Eprint
  {http://arxiv.org/abs/2008.11909} {arXiv:2008.11909 [hep-ph]} \BibitemShut
  {NoStop}%
\bibitem [{\citenamefont {Chun}\ and\ \citenamefont
  {Mondal}(2020)}]{Chun:2020uzw}%
  \BibitemOpen
  \bibfield  {author} {\bibinfo {author} {\bibfnamefont {J.~E.}\ \bibnamefont
  {Chun}}\ and\ \bibinfo {author} {\bibfnamefont {T.}~\bibnamefont {Mondal}},\
  }\href@noop {} {\  (\bibinfo {year} {2020})},\ \Eprint
  {http://arxiv.org/abs/2009.08314} {arXiv:2009.08314 [hep-ph]} \BibitemShut
  {NoStop}%
\bibitem [{\citenamefont {Chen}\ \emph {et~al.}(2020)\citenamefont {Chen},
  \citenamefont {Chiang},\ and\ \citenamefont {Yagyu}}]{Chen:2020tfr}%
  \BibitemOpen
  \bibfield  {author} {\bibinfo {author} {\bibfnamefont {K.-F.}\ \bibnamefont
  {Chen}}, \bibinfo {author} {\bibfnamefont {C.-W.}\ \bibnamefont {Chiang}}, \
  and\ \bibinfo {author} {\bibfnamefont {K.}~\bibnamefont {Yagyu}},\ }\href
  {\doibase 10.1007/JHEP09(2020)119} {\bibfield  {journal} {\bibinfo  {journal}
  {JHEP}\ }\textbf {\bibinfo {volume} {09}},\ \bibinfo {pages} {119} (\bibinfo
  {year} {2020})},\ \Eprint {http://arxiv.org/abs/2006.07929} {arXiv:2006.07929
  [hep-ph]} \BibitemShut {NoStop}%
\bibitem [{\citenamefont {Jana}\ \emph {et~al.}(2020)\citenamefont {Jana},
  \citenamefont {Vishnu}, \citenamefont {Rodejohann},\ and\ \citenamefont
  {Saad}}]{Jana:2020joi}%
  \BibitemOpen
  \bibfield  {author} {\bibinfo {author} {\bibfnamefont {S.}~\bibnamefont
  {Jana}}, \bibinfo {author} {\bibfnamefont {P.~K.}\ \bibnamefont {Vishnu}},
  \bibinfo {author} {\bibfnamefont {W.}~\bibnamefont {Rodejohann}}, \ and\
  \bibinfo {author} {\bibfnamefont {S.}~\bibnamefont {Saad}},\ }\href {\doibase
  10.1103/PhysRevD.102.075003} {\bibfield  {journal} {\bibinfo  {journal}
  {Phys. Rev. D}\ }\textbf {\bibinfo {volume} {102}},\ \bibinfo {pages}
  {075003} (\bibinfo {year} {2020})},\ \Eprint
  {http://arxiv.org/abs/2008.02377} {arXiv:2008.02377 [hep-ph]} \BibitemShut
  {NoStop}%
\bibitem [{\citenamefont {De~Jesus}\ \emph {et~al.}(2020)\citenamefont
  {De~Jesus}, \citenamefont {Kovalenko}, \citenamefont {Queiroz}, \citenamefont
  {Siqueira},\ and\ \citenamefont {Sinha}}]{deJesus:2020upp}%
  \BibitemOpen
  \bibfield  {author} {\bibinfo {author} {\bibfnamefont {A.~S.}\ \bibnamefont
  {De~Jesus}}, \bibinfo {author} {\bibfnamefont {S.}~\bibnamefont {Kovalenko}},
  \bibinfo {author} {\bibfnamefont {F.~S.}\ \bibnamefont {Queiroz}}, \bibinfo
  {author} {\bibfnamefont {C.}~\bibnamefont {Siqueira}}, \ and\ \bibinfo
  {author} {\bibfnamefont {K.}~\bibnamefont {Sinha}},\ }\href {\doibase
  10.1103/PhysRevD.102.035004} {\bibfield  {journal} {\bibinfo  {journal}
  {Phys. Rev. D}\ }\textbf {\bibinfo {volume} {102}},\ \bibinfo {pages}
  {035004} (\bibinfo {year} {2020})},\ \Eprint
  {http://arxiv.org/abs/2004.01200} {arXiv:2004.01200 [hep-ph]} \BibitemShut
  {NoStop}%
\bibitem [{\citenamefont {Zhou}\ and\ \citenamefont
  {Han}(2022)}]{newzhou:2022cql}%
  \BibitemOpen
  \bibfield  {author} {\bibinfo {author} {\bibfnamefont {Q.}~\bibnamefont
  {Zhou}}\ and\ \bibinfo {author} {\bibfnamefont {X.-F.}\ \bibnamefont {Han}},\
  }\href@noop {} {\  (\bibinfo {year} {2022})},\ \Eprint
  {http://arxiv.org/abs/2204.13027} {arXiv:2204.13027 [hep-ph]} \BibitemShut
  {NoStop}%
\bibitem [{\citenamefont {Iso}\ \emph {et~al.}(2009)\citenamefont {Iso},
  \citenamefont {Okada},\ and\ \citenamefont {Orikasa}}]{Iso:2009nw}%
  \BibitemOpen
  \bibfield  {author} {\bibinfo {author} {\bibfnamefont {S.}~\bibnamefont
  {Iso}}, \bibinfo {author} {\bibfnamefont {N.}~\bibnamefont {Okada}}, \ and\
  \bibinfo {author} {\bibfnamefont {Y.}~\bibnamefont {Orikasa}},\ }\href
  {\doibase 10.1103/PhysRevD.80.115007} {\bibfield  {journal} {\bibinfo
  {journal} {Phys. Rev. D}\ }\textbf {\bibinfo {volume} {80}},\ \bibinfo
  {pages} {115007} (\bibinfo {year} {2009})},\ \Eprint
  {http://arxiv.org/abs/0909.0128} {arXiv:0909.0128 [hep-ph]} \BibitemShut
  {NoStop}%
\bibitem [{\citenamefont {Bahrami}\ and\ \citenamefont
  {Frank}(2013)}]{Bahrami:2013bsa}%
  \BibitemOpen
  \bibfield  {author} {\bibinfo {author} {\bibfnamefont {S.}~\bibnamefont
  {Bahrami}}\ and\ \bibinfo {author} {\bibfnamefont {M.}~\bibnamefont
  {Frank}},\ }\href {\doibase 10.1103/PhysRevD.88.095002} {\bibfield  {journal}
  {\bibinfo  {journal} {Phys. Rev.}\ }\textbf {\bibinfo {volume} {D88}},\
  \bibinfo {pages} {095002} (\bibinfo {year} {2013})},\ \Eprint
  {http://arxiv.org/abs/1308.2847} {arXiv:1308.2847 [hep-ph]} \BibitemShut
  {NoStop}%
\bibitem [{\citenamefont {Bahrami}\ and\ \citenamefont
  {Frank}(2015)}]{Bahrami:2015mwa}%
  \BibitemOpen
  \bibfield  {author} {\bibinfo {author} {\bibfnamefont {S.}~\bibnamefont
  {Bahrami}}\ and\ \bibinfo {author} {\bibfnamefont {M.}~\bibnamefont
  {Frank}},\ }\href {\doibase 10.1103/PhysRevD.91.075003} {\bibfield  {journal}
  {\bibinfo  {journal} {Phys. Rev.}\ }\textbf {\bibinfo {volume} {D91}},\
  \bibinfo {pages} {075003} (\bibinfo {year} {2015})},\ \Eprint
  {http://arxiv.org/abs/1502.02680} {arXiv:1502.02680 [hep-ph]} \BibitemShut
  {NoStop}%
\bibitem [{\citenamefont {Bahrami}\ \emph {et~al.}(2017)\citenamefont
  {Bahrami}, \citenamefont {Frank}, \citenamefont {Ghosh}, \citenamefont
  {Ghosh},\ and\ \citenamefont {Saha}}]{Bahrami:2016has}%
  \BibitemOpen
  \bibfield  {author} {\bibinfo {author} {\bibfnamefont {S.}~\bibnamefont
  {Bahrami}}, \bibinfo {author} {\bibfnamefont {M.}~\bibnamefont {Frank}},
  \bibinfo {author} {\bibfnamefont {D.~K.}\ \bibnamefont {Ghosh}}, \bibinfo
  {author} {\bibfnamefont {N.}~\bibnamefont {Ghosh}}, \ and\ \bibinfo {author}
  {\bibfnamefont {I.}~\bibnamefont {Saha}},\ }\href {\doibase
  10.1103/PhysRevD.95.095024} {\bibfield  {journal} {\bibinfo  {journal} {Phys.
  Rev.}\ }\textbf {\bibinfo {volume} {D95}},\ \bibinfo {pages} {095024}
  (\bibinfo {year} {2017})},\ \Eprint {http://arxiv.org/abs/1612.06334}
  {arXiv:1612.06334 [hep-ph]} \BibitemShut {NoStop}%
\bibitem [{\citenamefont {Chakrabarty}\ \emph {et~al.}(2018)\citenamefont
  {Chakrabarty}, \citenamefont {Chiang}, \citenamefont {Ohata},\ and\
  \citenamefont {Tsumura}}]{Chakrabarty:2018qtt}%
  \BibitemOpen
  \bibfield  {author} {\bibinfo {author} {\bibfnamefont {N.}~\bibnamefont
  {Chakrabarty}}, \bibinfo {author} {\bibfnamefont {C.-W.}\ \bibnamefont
  {Chiang}}, \bibinfo {author} {\bibfnamefont {T.}~\bibnamefont {Ohata}}, \
  and\ \bibinfo {author} {\bibfnamefont {K.}~\bibnamefont {Tsumura}},\ }\href
  {\doibase 10.1007/JHEP12(2018)104} {\bibfield  {journal} {\bibinfo  {journal}
  {JHEP}\ }\textbf {\bibinfo {volume} {12}},\ \bibinfo {pages} {104} (\bibinfo
  {year} {2018})},\ \Eprint {http://arxiv.org/abs/1807.08167} {arXiv:1807.08167
  [hep-ph]} \BibitemShut {NoStop}%
\bibitem [{\citenamefont {Calibbi}\ and\ \citenamefont
  {Signorelli}(2018)}]{Calibbi:2017uvl}%
  \BibitemOpen
  \bibfield  {author} {\bibinfo {author} {\bibfnamefont {L.}~\bibnamefont
  {Calibbi}}\ and\ \bibinfo {author} {\bibfnamefont {G.}~\bibnamefont
  {Signorelli}},\ }\href {\doibase 10.1393/ncr/i2018-10144-0} {\bibfield
  {journal} {\bibinfo  {journal} {Riv. Nuovo Cim.}\ }\textbf {\bibinfo {volume}
  {41}},\ \bibinfo {pages} {71} (\bibinfo {year} {2018})},\ \Eprint
  {http://arxiv.org/abs/1709.00294} {arXiv:1709.00294 [hep-ph]} \BibitemShut
  {NoStop}%
\bibitem [{\citenamefont {Ishiwata}\ and\ \citenamefont
  {Wise}(2013)}]{Ishiwata:2013gma}%
  \BibitemOpen
  \bibfield  {author} {\bibinfo {author} {\bibfnamefont {K.}~\bibnamefont
  {Ishiwata}}\ and\ \bibinfo {author} {\bibfnamefont {M.~B.}\ \bibnamefont
  {Wise}},\ }\href {\doibase 10.1103/PhysRevD.88.055009} {\bibfield  {journal}
  {\bibinfo  {journal} {Phys. Rev. D}\ }\textbf {\bibinfo {volume} {88}},\
  \bibinfo {pages} {055009} (\bibinfo {year} {2013})},\ \Eprint
  {http://arxiv.org/abs/1307.1112} {arXiv:1307.1112 [hep-ph]} \BibitemShut
  {NoStop}%
\bibitem [{\citenamefont {Dermisek}\ \emph {et~al.}(2014)\citenamefont
  {Dermisek}, \citenamefont {Raval},\ and\ \citenamefont
  {Shin}}]{Dermisek:2014cia}%
  \BibitemOpen
  \bibfield  {author} {\bibinfo {author} {\bibfnamefont {R.}~\bibnamefont
  {Dermisek}}, \bibinfo {author} {\bibfnamefont {A.}~\bibnamefont {Raval}}, \
  and\ \bibinfo {author} {\bibfnamefont {S.}~\bibnamefont {Shin}},\ }\href
  {\doibase 10.1103/PhysRevD.90.034023} {\bibfield  {journal} {\bibinfo
  {journal} {Phys. Rev. D}\ }\textbf {\bibinfo {volume} {90}},\ \bibinfo
  {pages} {034023} (\bibinfo {year} {2014})},\ \Eprint
  {http://arxiv.org/abs/1406.7018} {arXiv:1406.7018 [hep-ph]} \BibitemShut
  {NoStop}%
\bibitem [{\citenamefont {Dicus}\ and\ \citenamefont
  {Mathur}(1973)}]{PhysRevD.7.3111}%
  \BibitemOpen
  \bibfield  {author} {\bibinfo {author} {\bibfnamefont {D.~A.}\ \bibnamefont
  {Dicus}}\ and\ \bibinfo {author} {\bibfnamefont {V.~S.}\ \bibnamefont
  {Mathur}},\ }\href {\doibase 10.1103/PhysRevD.7.3111} {\bibfield  {journal}
  {\bibinfo  {journal} {Phys. Rev. D}\ }\textbf {\bibinfo {volume} {7}},\
  \bibinfo {pages} {3111} (\bibinfo {year} {1973})}\BibitemShut {NoStop}%
\bibitem [{\citenamefont {Lee}\ \emph {et~al.}(1977)\citenamefont {Lee},
  \citenamefont {Quigg},\ and\ \citenamefont {Thacker}}]{PhysRevD.16.1519}%
  \BibitemOpen
  \bibfield  {author} {\bibinfo {author} {\bibfnamefont {B.~W.}\ \bibnamefont
  {Lee}}, \bibinfo {author} {\bibfnamefont {C.}~\bibnamefont {Quigg}}, \ and\
  \bibinfo {author} {\bibfnamefont {H.~B.}\ \bibnamefont {Thacker}},\ }\href
  {\doibase 10.1103/PhysRevD.16.1519} {\bibfield  {journal} {\bibinfo
  {journal} {Phys. Rev. D}\ }\textbf {\bibinfo {volume} {16}},\ \bibinfo
  {pages} {1519} (\bibinfo {year} {1977})}\BibitemShut {NoStop}%
\bibitem [{\citenamefont {Arhrib}\ \emph {et~al.}(2011)\citenamefont {Arhrib},
  \citenamefont {Benbrik}, \citenamefont {Chabab}, \citenamefont {Moultaka},
  \citenamefont {Peyran\`ere}, \citenamefont {Rahili},\ and\ \citenamefont
  {Ramadan}}]{PhysRevD.84.095005}%
  \BibitemOpen
  \bibfield  {author} {\bibinfo {author} {\bibfnamefont {A.}~\bibnamefont
  {Arhrib}}, \bibinfo {author} {\bibfnamefont {R.}~\bibnamefont {Benbrik}},
  \bibinfo {author} {\bibfnamefont {M.}~\bibnamefont {Chabab}}, \bibinfo
  {author} {\bibfnamefont {G.}~\bibnamefont {Moultaka}}, \bibinfo {author}
  {\bibfnamefont {M.~C.}\ \bibnamefont {Peyran\`ere}}, \bibinfo {author}
  {\bibfnamefont {L.}~\bibnamefont {Rahili}}, \ and\ \bibinfo {author}
  {\bibfnamefont {J.}~\bibnamefont {Ramadan}},\ }\href {\doibase
  10.1103/PhysRevD.84.095005} {\bibfield  {journal} {\bibinfo  {journal} {Phys.
  Rev. D}\ }\textbf {\bibinfo {volume} {84}},\ \bibinfo {pages} {095005}
  (\bibinfo {year} {2011})}\BibitemShut {NoStop}%
\bibitem [{\citenamefont {Patrignani}\ \emph {et~al.}(2016)\citenamefont
  {Patrignani} \emph {et~al.}}]{Patrignani:2016xqp}%
  \BibitemOpen
  \bibfield  {author} {\bibinfo {author} {\bibfnamefont {C.}~\bibnamefont
  {Patrignani}} \emph {et~al.} (\bibinfo {collaboration} {Particle Data
  Group}),\ }\href {\doibase 10.1088/1674-1137/40/10/100001} {\bibfield
  {journal} {\bibinfo  {journal} {Chin. Phys.}\ }\textbf {\bibinfo {volume}
  {C40}},\ \bibinfo {pages} {100001} (\bibinfo {year} {2016})}\BibitemShut
  {NoStop}%
\bibitem [{\citenamefont {Tanabashi}\ \emph {et~al.}(2018)\citenamefont
  {Tanabashi} \emph {et~al.}}]{PhysRevD.98.030001}%
  \BibitemOpen
  \bibfield  {author} {\bibinfo {author} {\bibfnamefont {M.}~\bibnamefont
  {Tanabashi}} \emph {et~al.} (\bibinfo {collaboration} {Particle Data
  Group}),\ }\href {\doibase 10.1103/PhysRevD.98.030001} {\bibfield  {journal}
  {\bibinfo  {journal} {Phys. Rev. D}\ }\textbf {\bibinfo {volume} {98}},\
  \bibinfo {pages} {030001} (\bibinfo {year} {2018})}\BibitemShut {NoStop}%
\bibitem [{ATL(2022)}]{ATLAS:2022yzd}%
  \BibitemOpen
  \href@noop {} {\  (\bibinfo {year} {2022})}\BibitemShut {NoStop}%
\bibitem [{CMS(2017)}]{CMS:2017pet}%
  \BibitemOpen
  \href@noop {} {\  (\bibinfo {year} {2017})}\BibitemShut {NoStop}%
\bibitem [{\citenamefont {Aad}\ \emph {et~al.}(2021)\citenamefont {Aad} \emph
  {et~al.}}]{ATLAS:2021jol}%
  \BibitemOpen
  \bibfield  {author} {\bibinfo {author} {\bibfnamefont {G.}~\bibnamefont
  {Aad}} \emph {et~al.} (\bibinfo {collaboration} {ATLAS}),\ }\href {\doibase
  10.1007/JHEP06(2021)146} {\bibfield  {journal} {\bibinfo  {journal} {JHEP}\
  }\textbf {\bibinfo {volume} {06}},\ \bibinfo {pages} {146} (\bibinfo {year}
  {2021})},\ \Eprint {http://arxiv.org/abs/2101.11961} {arXiv:2101.11961
  [hep-ex]} \BibitemShut {NoStop}%
\bibitem [{\citenamefont {Djouadi}(2008{\natexlab{a}})}]{Djouadi:2005gi}%
  \BibitemOpen
  \bibfield  {author} {\bibinfo {author} {\bibfnamefont {A.}~\bibnamefont
  {Djouadi}},\ }\href {\doibase 10.1016/j.physrep.2007.10.004} {\bibfield
  {journal} {\bibinfo  {journal} {Phys. Rept.}\ }\textbf {\bibinfo {volume}
  {457}},\ \bibinfo {pages} {1} (\bibinfo {year} {2008}{\natexlab{a}})},\
  \Eprint {http://arxiv.org/abs/hep-ph/0503172} {arXiv:hep-ph/0503172 [hep-ph]}
  \BibitemShut {NoStop}%
\bibitem [{\citenamefont {Djouadi}(2008{\natexlab{b}})}]{Djouadi:2005gj}%
  \BibitemOpen
  \bibfield  {author} {\bibinfo {author} {\bibfnamefont {A.}~\bibnamefont
  {Djouadi}},\ }\href {\doibase 10.1016/j.physrep.2007.10.005} {\bibfield
  {journal} {\bibinfo  {journal} {Phys. Rept.}\ }\textbf {\bibinfo {volume}
  {459}},\ \bibinfo {pages} {1} (\bibinfo {year} {2008}{\natexlab{b}})},\
  \Eprint {http://arxiv.org/abs/hep-ph/0503173} {arXiv:hep-ph/0503173 [hep-ph]}
  \BibitemShut {NoStop}%
\bibitem [{\citenamefont {Arhrib}\ \emph {et~al.}(2012)\citenamefont {Arhrib},
  \citenamefont {Benbrik}, \citenamefont {Chabab}, \citenamefont {Moultaka},\
  and\ \citenamefont {Rahili}}]{Arhrib:2011vc}%
  \BibitemOpen
  \bibfield  {author} {\bibinfo {author} {\bibfnamefont {A.}~\bibnamefont
  {Arhrib}}, \bibinfo {author} {\bibfnamefont {R.}~\bibnamefont {Benbrik}},
  \bibinfo {author} {\bibfnamefont {M.}~\bibnamefont {Chabab}}, \bibinfo
  {author} {\bibfnamefont {G.}~\bibnamefont {Moultaka}}, \ and\ \bibinfo
  {author} {\bibfnamefont {L.}~\bibnamefont {Rahili}},\ }\href {\doibase
  10.1007/JHEP04(2012)136} {\bibfield  {journal} {\bibinfo  {journal} {JHEP}\
  }\textbf {\bibinfo {volume} {04}},\ \bibinfo {pages} {136} (\bibinfo {year}
  {2012})},\ \Eprint {http://arxiv.org/abs/1112.5453} {arXiv:1112.5453
  [hep-ph]} \BibitemShut {NoStop}%
\bibitem [{\citenamefont {Aaboud}\ \emph {et~al.}(2018)\citenamefont {Aaboud}
  \emph {et~al.}}]{Aaboud:2018xdt}%
  \BibitemOpen
  \bibfield  {author} {\bibinfo {author} {\bibfnamefont {M.}~\bibnamefont
  {Aaboud}} \emph {et~al.} (\bibinfo {collaboration} {ATLAS}),\ }\href
  {\doibase 10.1103/PhysRevD.98.052005} {\bibfield  {journal} {\bibinfo
  {journal} {Phys. Rev.}\ }\textbf {\bibinfo {volume} {D98}},\ \bibinfo {pages}
  {052005} (\bibinfo {year} {2018})},\ \Eprint
  {http://arxiv.org/abs/1802.04146} {arXiv:1802.04146 [hep-ex]} \BibitemShut
  {NoStop}%
\bibitem [{\citenamefont {Sirunyan}\ \emph {et~al.}(2018)\citenamefont
  {Sirunyan} \emph {et~al.}}]{Sirunyan:2018ouh}%
  \BibitemOpen
  \bibfield  {author} {\bibinfo {author} {\bibfnamefont {A.~M.}\ \bibnamefont
  {Sirunyan}} \emph {et~al.} (\bibinfo {collaboration} {CMS}),\ }\href
  {\doibase 10.1007/JHEP11(2018)185} {\bibfield  {journal} {\bibinfo  {journal}
  {JHEP}\ }\textbf {\bibinfo {volume} {11}},\ \bibinfo {pages} {185} (\bibinfo
  {year} {2018})},\ \Eprint {http://arxiv.org/abs/1804.02716} {arXiv:1804.02716
  [hep-ex]} \BibitemShut {NoStop}%
\bibitem [{\citenamefont {Baldini}\ \emph {et~al.}(2016)\citenamefont {Baldini}
  \emph {et~al.}}]{TheMEG:2016wtm}%
  \BibitemOpen
  \bibfield  {author} {\bibinfo {author} {\bibfnamefont {A.~M.}\ \bibnamefont
  {Baldini}} \emph {et~al.} (\bibinfo {collaboration} {MEG}),\ }\href {\doibase
  10.1140/epjc/s10052-016-4271-x} {\bibfield  {journal} {\bibinfo  {journal}
  {Eur. Phys. J.}\ }\textbf {\bibinfo {volume} {C76}},\ \bibinfo {pages} {434}
  (\bibinfo {year} {2016})},\ \Eprint {http://arxiv.org/abs/1605.05081}
  {arXiv:1605.05081 [hep-ex]} \BibitemShut {NoStop}%
\bibitem [{\citenamefont {Aubert}\ \emph {et~al.}(2010)\citenamefont {Aubert}
  \emph {et~al.}}]{Aubert:2009ag}%
  \BibitemOpen
  \bibfield  {author} {\bibinfo {author} {\bibfnamefont {B.}~\bibnamefont
  {Aubert}} \emph {et~al.} (\bibinfo {collaboration} {BaBar}),\ }\href
  {\doibase 10.1103/PhysRevLett.104.021802} {\bibfield  {journal} {\bibinfo
  {journal} {Phys. Rev. Lett.}\ }\textbf {\bibinfo {volume} {104}},\ \bibinfo
  {pages} {021802} (\bibinfo {year} {2010})},\ \Eprint
  {http://arxiv.org/abs/0908.2381} {arXiv:0908.2381 [hep-ex]} \BibitemShut
  {NoStop}%
\bibitem [{\citenamefont {Bellgardt}\ \emph {et~al.}(1988)\citenamefont
  {Bellgardt} \emph {et~al.}}]{Bellgardt:1987du}%
  \BibitemOpen
  \bibfield  {author} {\bibinfo {author} {\bibfnamefont {U.}~\bibnamefont
  {Bellgardt}} \emph {et~al.} (\bibinfo {collaboration} {SINDRUM}),\ }\href
  {\doibase 10.1016/0550-3213(88)90462-2} {\bibfield  {journal} {\bibinfo
  {journal} {Nucl. Phys.}\ }\textbf {\bibinfo {volume} {B299}},\ \bibinfo
  {pages} {1} (\bibinfo {year} {1988})}\BibitemShut {NoStop}%
\bibitem [{\citenamefont {Amhis}\ \emph {et~al.}(2017)\citenamefont {Amhis}
  \emph {et~al.}}]{Amhis:2016xyh}%
  \BibitemOpen
  \bibfield  {author} {\bibinfo {author} {\bibfnamefont {Y.}~\bibnamefont
  {Amhis}} \emph {et~al.} (\bibinfo {collaboration} {HFLAV}),\ }\href {\doibase
  10.1140/epjc/s10052-017-5058-4} {\bibfield  {journal} {\bibinfo  {journal}
  {Eur. Phys. J.}\ }\textbf {\bibinfo {volume} {C77}},\ \bibinfo {pages} {895}
  (\bibinfo {year} {2017})},\ \Eprint {http://arxiv.org/abs/1612.07233}
  {arXiv:1612.07233 [hep-ex]} \BibitemShut {NoStop}%
\bibitem [{\citenamefont {Anastasiou}\ \emph {et~al.}(2009)\citenamefont
  {Anastasiou}, \citenamefont {Furlan},\ and\ \citenamefont
  {Santiago}}]{Anastasiou:2009rv}%
  \BibitemOpen
  \bibfield  {author} {\bibinfo {author} {\bibfnamefont {C.}~\bibnamefont
  {Anastasiou}}, \bibinfo {author} {\bibfnamefont {E.}~\bibnamefont {Furlan}},
  \ and\ \bibinfo {author} {\bibfnamefont {J.}~\bibnamefont {Santiago}},\
  }\href {\doibase 10.1103/PhysRevD.79.075003} {\bibfield  {journal} {\bibinfo
  {journal} {Phys. Rev. D}\ }\textbf {\bibinfo {volume} {79}},\ \bibinfo
  {pages} {075003} (\bibinfo {year} {2009})},\ \Eprint
  {http://arxiv.org/abs/0901.2117} {arXiv:0901.2117 [hep-ph]} \BibitemShut
  {NoStop}%
\bibitem [{\citenamefont {Lavoura}\ and\ \citenamefont
  {Silva}(1993)}]{PhysRevD.47.2046}%
  \BibitemOpen
  \bibfield  {author} {\bibinfo {author} {\bibfnamefont {L.}~\bibnamefont
  {Lavoura}}\ and\ \bibinfo {author} {\bibfnamefont {J.~a.~P.}\ \bibnamefont
  {Silva}},\ }\href {\doibase 10.1103/PhysRevD.47.2046} {\bibfield  {journal}
  {\bibinfo  {journal} {Phys. Rev. D}\ }\textbf {\bibinfo {volume} {47}},\
  \bibinfo {pages} {2046} (\bibinfo {year} {1993})}\BibitemShut {NoStop}%
\bibitem [{\citenamefont {Chen}\ and\ \citenamefont
  {Dawson}(2004)}]{PhysRevD.70.015003}%
  \BibitemOpen
  \bibfield  {author} {\bibinfo {author} {\bibfnamefont {M.-C.}\ \bibnamefont
  {Chen}}\ and\ \bibinfo {author} {\bibfnamefont {S.}~\bibnamefont {Dawson}},\
  }\href {\doibase 10.1103/PhysRevD.70.015003} {\bibfield  {journal} {\bibinfo
  {journal} {Phys. Rev. D}\ }\textbf {\bibinfo {volume} {70}},\ \bibinfo
  {pages} {015003} (\bibinfo {year} {2004})}\BibitemShut {NoStop}%
\bibitem [{\citenamefont {Chen}\ \emph {et~al.}(2017)\citenamefont {Chen},
  \citenamefont {Dawson},\ and\ \citenamefont {Furlan}}]{Chen:2017hak}%
  \BibitemOpen
  \bibfield  {author} {\bibinfo {author} {\bibfnamefont {C.-Y.}\ \bibnamefont
  {Chen}}, \bibinfo {author} {\bibfnamefont {S.}~\bibnamefont {Dawson}}, \ and\
  \bibinfo {author} {\bibfnamefont {E.}~\bibnamefont {Furlan}},\ }\href
  {\doibase 10.1103/PhysRevD.96.015006} {\bibfield  {journal} {\bibinfo
  {journal} {Phys. Rev. D}\ }\textbf {\bibinfo {volume} {96}},\ \bibinfo
  {pages} {015006} (\bibinfo {year} {2017})},\ \Eprint
  {http://arxiv.org/abs/1703.06134} {arXiv:1703.06134 [hep-ph]} \BibitemShut
  {NoStop}%
\bibitem [{\citenamefont {Lavoura}\ and\ \citenamefont
  {Li}(1994)}]{Lavoura:1993nq}%
  \BibitemOpen
  \bibfield  {author} {\bibinfo {author} {\bibfnamefont {L.}~\bibnamefont
  {Lavoura}}\ and\ \bibinfo {author} {\bibfnamefont {L.-F.}\ \bibnamefont
  {Li}},\ }\href {\doibase 10.1103/PhysRevD.49.1409} {\bibfield  {journal}
  {\bibinfo  {journal} {Phys. Rev. D}\ }\textbf {\bibinfo {volume} {49}},\
  \bibinfo {pages} {1409} (\bibinfo {year} {1994})},\ \Eprint
  {http://arxiv.org/abs/hep-ph/9309262} {arXiv:hep-ph/9309262} \BibitemShut
  {NoStop}%
\bibitem [{\citenamefont {Passarino}\ and\ \citenamefont
  {Veltman}(1979)}]{PASSARINO1979151}%
  \BibitemOpen
  \bibfield  {author} {\bibinfo {author} {\bibfnamefont {G.}~\bibnamefont
  {Passarino}}\ and\ \bibinfo {author} {\bibfnamefont {M.}~\bibnamefont
  {Veltman}},\ }\href {\doibase https://doi.org/10.1016/0550-3213(79)90234-7}
  {\bibfield  {journal} {\bibinfo  {journal} {Nuclear Physics B}\ }\textbf
  {\bibinfo {volume} {160}},\ \bibinfo {pages} {151} (\bibinfo {year}
  {1979})}\BibitemShut {NoStop}%
\bibitem [{\citenamefont {Hahn}\ and\ \citenamefont
  {Pérez-Victoria}(1999)}]{HAHN1999153}%
  \BibitemOpen
  \bibfield  {author} {\bibinfo {author} {\bibfnamefont {T.}~\bibnamefont
  {Hahn}}\ and\ \bibinfo {author} {\bibfnamefont {M.}~\bibnamefont
  {Pérez-Victoria}},\ }\href {\doibase
  https://doi.org/10.1016/S0010-4655(98)00173-8} {\bibfield  {journal}
  {\bibinfo  {journal} {Computer Physics Communications}\ }\textbf {\bibinfo
  {volume} {118}},\ \bibinfo {pages} {153} (\bibinfo {year}
  {1999})}\BibitemShut {NoStop}%
\bibitem [{\citenamefont {Lindner}\ \emph {et~al.}(2018)\citenamefont
  {Lindner}, \citenamefont {Platscher},\ and\ \citenamefont
  {Queiroz}}]{Lindner:2016bgg}%
  \BibitemOpen
  \bibfield  {author} {\bibinfo {author} {\bibfnamefont {M.}~\bibnamefont
  {Lindner}}, \bibinfo {author} {\bibfnamefont {M.}~\bibnamefont {Platscher}},
  \ and\ \bibinfo {author} {\bibfnamefont {F.~S.}\ \bibnamefont {Queiroz}},\
  }\href {\doibase 10.1016/j.physrep.2017.12.001} {\bibfield  {journal}
  {\bibinfo  {journal} {Phys. Rept.}\ }\textbf {\bibinfo {volume} {731}},\
  \bibinfo {pages} {1} (\bibinfo {year} {2018})},\ \Eprint
  {http://arxiv.org/abs/1610.06587} {arXiv:1610.06587 [hep-ph]} \BibitemShut
  {NoStop}%
\bibitem [{\citenamefont {Akeroyd}\ \emph {et~al.}(2009)\citenamefont
  {Akeroyd}, \citenamefont {Aoki},\ and\ \citenamefont
  {Sugiyama}}]{Akeroyd:2009nu}%
  \BibitemOpen
  \bibfield  {author} {\bibinfo {author} {\bibfnamefont {A.}~\bibnamefont
  {Akeroyd}}, \bibinfo {author} {\bibfnamefont {M.}~\bibnamefont {Aoki}}, \
  and\ \bibinfo {author} {\bibfnamefont {H.}~\bibnamefont {Sugiyama}},\ }\href
  {\doibase 10.1103/PhysRevD.79.113010} {\bibfield  {journal} {\bibinfo
  {journal} {Phys. Rev. D}\ }\textbf {\bibinfo {volume} {79}},\ \bibinfo
  {pages} {113010} (\bibinfo {year} {2009})},\ \Eprint
  {http://arxiv.org/abs/0904.3640} {arXiv:0904.3640 [hep-ph]} \BibitemShut
  {NoStop}%
\bibitem [{\citenamefont {Akeroyd}\ and\ \citenamefont
  {Chiang}(2009)}]{Akeroyd:2009hb}%
  \BibitemOpen
  \bibfield  {author} {\bibinfo {author} {\bibfnamefont {A.~G.}\ \bibnamefont
  {Akeroyd}}\ and\ \bibinfo {author} {\bibfnamefont {C.-W.}\ \bibnamefont
  {Chiang}},\ }\href {\doibase 10.1103/PhysRevD.80.113010} {\bibfield
  {journal} {\bibinfo  {journal} {Phys. Rev.}\ }\textbf {\bibinfo {volume}
  {D80}},\ \bibinfo {pages} {113010} (\bibinfo {year} {2009})},\ \Eprint
  {http://arxiv.org/abs/0909.4419} {arXiv:0909.4419 [hep-ph]} \BibitemShut
  {NoStop}%
\end{thebibliography}%
\end{document}